\documentclass[12pt]{article}
\usepackage{amsmath}
\usepackage{graphicx}
\usepackage{natbib}
\usepackage{url}
\usepackage{mathtools}
\usepackage{amsfonts}
\usepackage{amssymb}
\usepackage{booktabs}
\usepackage[tableposition=top]{caption}
\usepackage{bm}
\usepackage[nodisplayskipstretch]{setspace}
\usepackage{float}
\usepackage{placeins}
\usepackage{color}
\usepackage[colorlinks=true,linkcolor=blue,citecolor=black]{hyperref}

\usepackage{amsthm}
\theoremstyle{plain}
\newtheorem{theorem}{Theorem}
\newtheorem{lemma}{Lemma}
\newtheorem{proposition}{Proposition}
\newtheorem{corollary}{Corollary}
\theoremstyle{definition}
\newtheorem{definition}{Definition}
\newtheorem{assumption}{Assumption}

\newcommand{\blind}{1}

\allowdisplaybreaks

\newcommand{\bbP}{\mathbb{P}}
\newcommand{\bbE}{\mathbb{E}}
\newcommand{\bbR}{\mathbb{R}}
\newcommand{\cZ}{\mathcal{Z}}
\newcommand{\cC}{\mathcal{C}}
\newcommand{\cD}{\mathcal{D}}
\newcommand{\cQ}{\mathcal{Q}}
\newcommand{\Yhat}{\widehat{Y}}
\newcommand{\tauS}{\widehat\tau_{\rm S}}
\newcommand{\tauW}{\widehat\tau_{\rm W}}
\newcommand{\ttil}{\widetilde\tau}
\newcommand{\abar}{\overline{a}}

\newcommand{\vhat}{\widehat v}
\newcommand{\Ystar}{Y^{*}}
\newcommand{\astar}{a^{*}}
\newcommand{\abarstar}{\overline{a}^{\,*}}
\newcommand{\ahstar}{a^{*}_{\rm h}}
\newcommand{\F}{\mathbb{F}_2}
\numberwithin{equation}{section}
\newcommand{\hd}[2]{\begin{tabular}[b]{@{}c@{}}#1\\#2\end{tabular}}

\begin{document}

\def\spacingset#1{\renewcommand{\baselinestretch}%
{#1}\small\normalsize} \spacingset{1}


\if1\blind
{
  \title{\Large \bf The precision of fractional factorial designs under randomization}
  \author{Xinyuan Chen\vspace{0.2cm}\\
    Department of Mathematics and Statistics, \\
    Mississippi State University, Mississippi State, MS, USA \\
    xchen@math.msstate.edu}
  \date{}
  \maketitle
} \fi

\if0\blind
{
  \bigskip
  \bigskip
  \bigskip
  \begin{center}
    {\Large \bf The Precision of Fractional Factorial Designs under Randomization}
\end{center}
  \medskip
} \fi

\begin{abstract}
    A regular fraction of a two-level factorial design runs fewer treatment combinations (cells), with more units in each. With the number of units fixed, more units per cell do not by themselves reduce the variance. Averaged over the fractions of one defining contrast subgroup, a fraction's variance is smaller than the full design's by the unit-to-unit variation of the aliased effects, a gain the usual variance estimator misses. Assuming the aliased effects are zero, we propose a full-design estimator that weights the fractions' contrasts by inverse variance. Its estimated variance never exceeds the saturated estimator's. In large samples, it is proportional to the harmonic mean of the fractions' average cell variances, and a fraction's, on average, to the arithmetic mean. We test this assumption, but choosing between the two estimators by the test can lose coverage. The results extend to unequal allocations, and the large-sample results to covariate adjustment.
\end{abstract}

\noindent%
{\it Keywords:} design-based inference; factorial design; fractional factorial design; randomized experiments.

\spacingset{1.75} 

\section{Introduction}

A $2^K$ factorial experiment assigns units to all combinations of $K$ two-level factors. When $2^K$ is large, experimenters often run a regular fraction, which uses $2^{K-p}$ of the treatment combinations (cells) \citep[\S5]{box1961fractionalI, box1961fractionalII, wu2021experiments}. With the same number of units, each cell in the fraction receives $2^p$ times as many units as a cell of the full design. However, this comes with the cost of aliasing. That is, each factorial effect outside the defining contrast subgroup is confounded with $2^p-1$ others, and the fraction estimates their signed sum. The usual justification is the effect hierarchy principle, under which the aliased interactions are negligible \citep{wu2021experiments, pashley2023causal}.

We study whether running a fraction improves precision in the design-based framework, in which the potential outcomes and covariates are fixed, and the assignment is the sole source of randomness. \citet*{dasgupta2015causal} developed this framework for $2^K$ designs, and \citet{pashley2023causal} extended it to fractions. Under additivity, i.e., when every factorial effect is the same for every unit, \citet{pashley2023causal} found that a fraction and the full design with the same number of units have the same variance. \citet{zhao2023covariate} compared a fraction with a full design, analyzed by a factor-unsaturated regression, via simulation, and left the theory to future research. Equal precision with a fixed number of units is also the classical conclusion under a linear model with constant error variance \citep*[e.g.,][]{collins2009design}.

This article gives the comparison without additivity, for regular fractions and balanced designs with the same number of units. The extra units per cell do not reduce the variance by themselves. Averaged over the $2^p$ fractions with the same defining contrast subgroup, the variance of a fraction is smaller than that of the full design by the unit-to-unit variation of the aliased effects. The usual variance estimator does not detect this gain, and in our simulation, neither does the sharp variance bound of \citet*{aronow2014sharp}, when extended to a coset (\S\ref{supp:sim-sharp} of the Supplementary Material). Under the fraction's assumption that the aliased effects are zero, we propose a weighted estimator under the full design, which combines the contrasts of these fractions with inverse-variance weights. In large samples, its estimated variance is proportional to the harmonic mean of the average cell variances of the fractions, while that of a fraction, on average, is proportional to their arithmetic mean. The estimated variance of the weighted estimator is also never larger than that of the saturated estimator, which needs no assumption on the aliased effects.

Under the full design, we also provide a test for the fraction's assumption. However, choosing between the weighted and saturated estimators by this test can lose coverage, as is common after a pretest \citep{leeb2005model}. In large samples, the results extend to covariate adjustment by the fully interacted regression of \citet{lin2013agnostic}. The exact and large-sample results, and the test, extend to unequal allocations, for which \citet{ravichandran2024optimal} derived the optimal allocations in $2^K$ designs. The comparisons with the fractions then hold when each cell of a fraction has $2^p$ times as many units as the same cell of the full design. When the potential outcomes do not depend on the stage, all results also hold when the cosets of the full design are run in separate stages planned in advance. A fold-over planned in advance is then the full design when $p=1$, and a fraction with a smaller defining contrast subgroup when $p\geq2$ (\S\ref{supp:unequal} of the Supplementary Material).

\section{Setup}

\subsection{Factorial effects}\label{sec:effects}

Consider $N$ units and $K$ factors with levels $-1$ and $+1$. A treatment combination (cell) is $z=(z_1,\ldots,z_K)\in\cZ\equiv\{-1,+1\}^K$, and unit $i$ has potential outcome $Y_i(z)$. Write $\overline Y(z)=N^{-1}\sum_{i=1}^N Y_i(z)$, and assume the stable unit treatment value assumption (SUTVA) holds throughout. A factorial effect is labeled by a subset $k$ of $\{1,\ldots,K\}$, written as a word, e.g., $23$ stands for $\{2,3\}$, with the empty word being $\emptyset$. For two words $k$ and $w$, the product $kw$ denotes their symmetric difference. The contrast of word $k$ is $g_k(z)=\prod_{l\in k}z_l$, with $g_\emptyset(z)=1$, so that $g_k(z)g_w(z)=g_{kw}(z)$. Following \citet*{dasgupta2015causal}, the factorial effects of word $k\neq\emptyset$ for unit $i$ and for the population respectively are
\begin{equation}\label{eq:effect}
    \tau_i(k)=\frac{1}{2^{K-1}}\sum_{z\in\cZ}g_k(z)Y_i(z),\qquad \tau(k)=\frac{1}{N}\sum_{i=1}^N\tau_i(k).
\end{equation}
For unit-level quantities $b_1,\ldots,b_N$, write $S^2(b)=(N-1)^{-1}\sum_{i=1}^N(b_i-\overline b)^2$ for their finite-population variance. Thus, $S^2\{\tau(k)\}$ is the variance of $\tau_1(k),\ldots,\tau_N(k)$, and the cell variance $S^2(z)\equiv S^2\{Y(z)\}$ is that of $Y_1(z),\ldots,Y_N(z)$.

\subsection{Regular fractions}\label{sec:fractions}

A regular fraction is built from $p$ generator words $w_1,\ldots,w_p$ such that no product of a nonempty subset of them equals $\emptyset$. The defining contrast subgroup $G$ is the set of $J=2^p$ products of subsets of the generators. Each sign vector $t\in\{-1,+1\}^p$ defines a coset $\{z\in\cZ: g_{w_r}(z)=t_r,\ r=1,\ldots,p\}$. Denote the $J$ cosets by $\cC_1,\ldots,\cC_J$, with $\cC_1$ given by $t=(1,\ldots,1)$. Each coset has $2^{K-p}$ cells, and the $J$ cosets partition $\cZ$. For $w\in G$, the contrast $g_w$ is constant on $\cC_j$, and denote its value by $\chi_j(w)$. A fraction runs the cells of one coset. For example, with $K=3$ and the generator $123$, the two cosets are $\cC_1=\{z:z_1z_2z_3=1\}$ and $\cC_2=\{z:z_1z_2z_3=-1\}$. Fix a target word $k\notin G$. For $w\in G\setminus\{\emptyset\}$, the word $kw$ is nonempty because $k\notin G$, and $g_{kw}(z)=\chi_j(w)\,g_k(z)$ for $z\in\cC_j$. Thus, on the cells of one coset, the contrasts of $k$ and $kw$ agree up to sign. We therefore call $\tau_i(kw)$ and $\tau(kw)$, for $w\in G\setminus\{\emptyset\}$, the aliased effects of unit $i$ and of the population. The coset contrast of unit $i$ and its population average respectively are
\begin{equation*}
    \ttil_{j,i}(k)=\frac{1}{2^{K-p-1}}\sum_{z\in\cC_j}g_k(z)Y_i(z),\qquad \ttil_j(k)=\frac{1}{N}\sum_{i=1}^N\ttil_{j,i}(k).
\end{equation*}

\begin{lemma}\label{lem:alias}
    For every unit $i$ and every coset $j$,
    \begin{equation} \label{eq:alias}
        \ttil_{j,i}(k)=\tau_i(k)+\sum_{w\in G\setminus\{\emptyset\}}\chi_j(w)\,\tau_i(kw).
    \end{equation}
\end{lemma}

\S\ref{supp:algebra} of the Supplementary Material proves Lemma \ref{lem:alias}. The coset contrast is the target effect plus the $J-1$ aliased effects $\tau_i(kw)$, each with the sign $\chi_j(w)$.

\subsection{Designs and estimators}\label{sec:designs}

\begin{definition}[Designs]\label{def:design}
    Let $N/2^K\geq2$ be an integer. The full design $\cD$ assigns the $N$ units by complete randomization to the $2^K$ cells of $\cZ$, with $N/2^K$ units in each cell. For $j=1,\ldots,J$, the fractional design $\cD_j$ assigns the $N$ units by complete randomization to the $2^{K-p}$ cells of $\cC_j$, with $N/2^{K-p}$ units in each cell.
\end{definition}

Running the cosets of $\cD$ in separate stages, after splitting the units completely at random, gives the same distribution of assignments as $\cD$. Therefore, the results below also hold for staged designs when the potential outcomes do not depend on the stage, including a fold-over planned in advance when $p=1$. For $p\geq2$, a fold-over is a fraction whose defining contrast subgroup has $2^{p-1}$ words, and the results for fractions extend (\S\ref{supp:staged} of the Supplementary Material).

Each estimator below is built from $\Yhat(z)$, an estimate of $\overline Y(z)$, and $\widehat\psi(z)$, an estimate of its variance, for every cell $z$ that the design runs. Let $n_z$ be the number of units in cell $z$ under the design at hand. The unadjusted version takes $\Yhat(z)$ to be the mean of the observed outcomes in cell $z$, and $\widehat\psi(z)=s^2(z)/n_z$ with $s^2(z)$ being their sample variance. The adjusted version uses covariates $x_i\in\bbR^d$ with mean $\overline x=N^{-1}\sum_{i=1}^N x_i$, where $\Yhat(z)$ is the intercept of the least squares fit of the observed outcome on $(1,x_i-\overline x)$ among the units in cell $z$, and $\widehat\psi(z)$ is its Eicker--Huber--White (EHW) variance estimate. When a within-cell fit is rank-deficient, $\Yhat(z)$ and $\widehat\psi(z)$ are those of the unadjusted version. Under $\cD$, each fit uses $N/2^K$ units, so the adjusted version needs $N/2^K$ to be large relative to $d$. This is the fully interacted adjustment of \citet{lin2013agnostic}, in the multiarmed form of \citet{zhao2023covariate}.

\begin{definition}[Estimators]\label{def:est}
    Under $\cD_j$, the fraction estimator and its variance estimator respectively are
    \begin{equation} \label{eq:fraction-formulas}
        \widehat\tau_j=\frac{1}{2^{K-p-1}}\sum_{z\in\cC_j}g_k(z)\Yhat(z),\qquad \vhat_j=\frac{1}{4^{K-p-1}}\sum_{z\in\cC_j}\widehat\psi(z).
    \end{equation}
    Under $\cD$, the saturated estimator and its variance estimator respectively are
    \begin{equation} \label{eq:saturated-formulas}
        \tauS=\frac{1}{2^{K-1}}\sum_{z\in\cZ}g_k(z)\Yhat(z),\qquad \vhat_{\rm S}=\frac{1}{4^{K-1}}\sum_{z\in\cZ}\widehat\psi(z).
    \end{equation}
    Under $\cD$, let $\widehat\delta_j$ and $\widehat u_j$ be the right-hand sides of \eqref{eq:fraction-formulas} computed on the cells of $\cC_j$. When every $\widehat u_j>0$, let $\widehat\lambda_j=\widehat u_j^{-1}/\sum_{j'=1}^J\widehat u_{j'}^{-1}$, and otherwise let $\widehat\lambda_j=1/J$. The weighted estimator and its variance estimator respectively are
    \begin{equation} \label{eq:weighted-formulas}
        \tauW=\sum_{j=1}^J\widehat\lambda_j\widehat\delta_j,\qquad \vhat_{\rm W}=\sum_{j=1}^J\widehat\lambda_j^2\,\widehat u_j.
    \end{equation}
    When every $\widehat u_j>0$, $\vhat_{\rm W}=(\sum_{j=1}^J\widehat u_j^{-1})^{-1}$.
\end{definition}

Each $\widehat\delta_j$ estimates the coset contrast $\ttil_j(k)$ from the full design. The saturated estimator is their equally weighted average, i.e., $\tauS=J^{-1}\sum_{j}\widehat\delta_j$. Averaging \eqref{eq:alias} over units gives $\ttil_j(k)=\sum_{w\in G}\chi_j(w)\tau(kw)$. Thus, when $\tau(kw)=0$ for every $w\neq\emptyset$, every $\widehat\delta_j$ estimates $\tau(k)$. When every $\widehat\psi(z)>0$, $\tauW$ is the feasible generalized least squares estimator under this restriction. Specifically, it equals $2^{-(K-1)}\sum_zg_k(z)\widetilde\mu(z)$, where $\widetilde\mu$ minimizes $\sum_z\{\Yhat(z)-\mu(z)\}^2/\widehat\psi(z)$ subject to $\sum_zg_{kw}(z)\mu(z)=0$ for every $w\neq\emptyset$, and $\vhat_{\rm W}$ is its generalized least squares variance (Lemma \ref{lem:s-gls} of the Supplementary Material). With $Z_i$ being the cell of unit $i$, the least squares regression of the observed outcome on an intercept and contrasts $g_m(Z_i)$, for $m$ in a set of distinct nonempty words that contains $k$, has coefficient $\tauS/2$ on $g_k(Z_i)$, where $\tauS$ is the unadjusted version, since these regressors are orthogonal in a balanced design (\S\ref{supp:ols} of the Supplementary Material). The factor-unsaturated specifications of \citet[\S5.4]{zhao2023covariate} without covariates are of this form. Their point estimates are therefore $\tauS$, although their EHW standard errors in general differ from $\vhat_{\rm S}^{1/2}$.

\section{Exact variances}\label{sec:exact}

This section concerns the unadjusted version. Expectations and variances are over the assignment, with the potential outcomes held fixed. Write $\mathrm{var}_j,\bbE_j$ under $\cD_j$, and $\mathrm{var},\bbE$ under $\cD$.

\begin{proposition}\label{prop:bias}
    Under Definitions \ref{def:design} and \ref{def:est} with the unadjusted version, for every coset $j$,
    \begin{equation}\label{eq:bias}
        \bbE_j(\widehat\tau_j)=\tau(k)+\sum_{w\in G\setminus\{\emptyset\}}\chi_j(w)\,\tau(kw),\qquad \bbE(\tauS)=\tau(k).
    \end{equation}
\end{proposition}

\S\ref{supp:exact} of the Supplementary Material proves Proposition \ref{prop:bias}. The first identity in \eqref{eq:bias} extends Result 1 of \citet{pashley2023causal} to every word $k\notin G$. A fraction estimates the signed sum of the target and its aliased effects, and the bias is zero for every coset only when every aliased effect is zero. In contrast, the second identity in \eqref{eq:bias} shows that the saturated estimator is unbiased for every set of potential outcomes.

Let $a_j=2^{-(K-p)}\sum_{z\in\cC_j}S^2(z)$ be the average cell variance on coset $j$, and $\abar=J^{-1}\sum_{j=1}^Ja_j$ be the average over all cells. Let $U$ be drawn uniformly from $\{1,\ldots,J\}$ and, given $U=j$, let the assignment be drawn from $\cD_j$. Let $\widehat\tau_U$ be the resulting fraction estimator, and write $\bbE_U$ and $\mathrm{var}_U$ for the expectation and variance over both $U$ and the assignment.

\begin{theorem}\label{thm:exact}
    Under Definitions \ref{def:design} and \ref{def:est} with the unadjusted version, the following hold for every set of potential outcomes.

    (i) For every coset $j$, $N\mathrm{var}_j(\widehat\tau_j)=4a_j-S^2\{\ttil_j(k)\}$, and $N\mathrm{var}(\tauS)=4\abar-S^2\{\tau(k)\}$.

    (ii) The average over cosets satisfies
    \begin{equation}\label{eq:identity}
        \frac{1}{J}\sum_{j=1}^J N\mathrm{var}_j(\widehat\tau_j)=N\mathrm{var}(\tauS)-\sum_{w\in G\setminus\{\emptyset\}}S^2\{\tau(kw)\}.
    \end{equation}

    (iii) For every coset $j$, $N \bbE_j(\vhat_j)=4a_j$, and $N\bbE(\vhat_{\rm S})=4\abar$. Hence $J^{-1}\sum_{j=1}^J N\bbE_j(\vhat_j)=N\bbE(\vhat_{\rm S})$.

    (iv) $\bbE_U(\widehat\tau_U)=\tau(k)$, and $N\mathrm{var}_U(\widehat\tau_U)=J^{-1}\sum_{j=1}^JN\mathrm{var}_j(\widehat\tau_j)+N\sum_{w\in G\setminus\{\emptyset\}}\tau(kw)^2$.
\end{theorem}

\S\ref{supp:exact} of the Supplementary Material proves Theorem \ref{thm:exact}. Part (i) for a fraction extends Result 2 of \citet{pashley2023causal} to every word $k\notin G$, and part (iii) extends their Result 4. The variance for the full design is from Theorem 2 of \citet*{dasgupta2015causal}. Part (i) shows that more units per cell do not reduce the variance on their own. The estimator of a fraction puts $2^p$ times as much weight on each cell, so a cell with $2^p$ times as many units contributes $4^p/2^p=2^p$ times as much variance. The fraction has $2^{-p}$ times as many cells, so both designs end at four times an average cell variance. Part (ii) compares the two designs on average over the coset that is run. The fraction, on average, has a variance no larger than that of the full design, and the difference reflects variation in the aliased effects across units.

When no effect varies across units, as in the comparison given by \citet{pashley2023causal}, every term $S^2\{\cdot\}$ is zero, and all cells have the same variance. Part (i) then gives the same variance for every coset and for the full design. Part (iii) shows that the variance estimators do not detect the gain in part (ii). Averaged over cosets, their expectations are the same under the two designs. Therefore, with the number of units fixed and no prior knowledge of which coset has the smaller cell variances, a fraction gives no gain in estimated precision on average. Part (iv) concerns the fraction run on a coset drawn at random. This randomized fraction is unbiased for $\tau(k)$. When $\tau(kw)=0$ for every $w\neq\emptyset$, $N$ times the variance of the randomized fraction equals the left-hand side of \eqref{eq:identity}, so \eqref{eq:identity} also compares the full design with the randomized fraction. Otherwise, $N$ times the variance of the randomized fraction exceeds the left-hand side of \eqref{eq:identity} by $N\sum_{w\neq\emptyset}\tau(kw)^2$.

\section{Inference and the weighted estimator}\label{sec:asy}

\subsection{Assumptions}\label{sec:assumptions}

Consider a sequence of finite populations with $N\to\infty$ through values allowed by Definition \ref{def:design}, and with $K$, $p$ and $d$ fixed. For the unadjusted version, let $\Ystar_i(z)=Y_i(z)$. For the adjusted version, let $\Ystar_i(z)=Y_i(z)-(x_i-\overline x)^\top\beta(z)$ with $\beta(z)=S_{xx}^{-1}S_{xY(z)}$, where $S_{xx}$ is the finite-population covariance matrix of $x$ and $S_{xY(z)}$ is the finite-population covariance of $x$ and $Y(z)$. Since $x$ is centered, $\Ystar$ and $Y$ have the same cell means, and therefore, every factorial effect $\tau(m)$, and the targets $\tau(k)$ and $\ttil_j(k)$, are the same for $\Ystar$ and $Y$. Write $\tau^*_i(m)$ and $\ttil^*_{j,i}(k)$ for the factorial effects and coset contrasts of $\Ystar$, $\tau^*(m)$ and $\ttil^*_j(k)$ for their respective population averages, and define
\begin{gather*}
    \astar_j=\frac{1}{2^{K-p}}\sum_{z\in\cC_j}S^2\{\Ystar(z)\},\quad \abarstar=\frac{1}{J}\sum_{j=1}^J\astar_j,\quad \ahstar=\frac{J}{\sum_{j=1}^J(\astar_j)^{-1}},\quad \lambda_j=\frac{(\astar_j)^{-1}}{\sum_{j'=1}^J(\astar_{j'})^{-1}}.
\end{gather*}
We call $\astar_j$ the average cell variance of coset $j$. For the unadjusted version, $\astar_j=a_j$.

\begin{assumption}\label{ass:maxdev}
    As $N\to\infty$, $\max_{z\in\cZ}\max_{1\leq i\leq N}\{Y_i(z)-\overline Y(z)\}^2/N\to0$.
\end{assumption}

\begin{assumption}\label{ass:limits}
    For all $z,z'\in\cZ$, the finite-population covariance of $Y(z)$ and $Y(z')$ has a finite limit, and the limit of $a_j$ is positive for every $j$.
\end{assumption}

\begin{assumption}\label{ass:cov}
    As $N\to\infty$, (i) the first two finite-population moments of $\{Y_i(z),x_i,x_iY_i(z):z\in\cZ\}$ have finite limits, (ii) $S_{xx}$ and its limit are nonsingular, (iii) $N^{-1}\sum_{i}Y_i(z)^4$, $N^{-1}\sum_i\|x_i\|^4$ and $N^{-1}\sum_i\|x_iY_i(z)\|^4$ are bounded for every $z$, and (iv) the limit of $\astar_j$ is positive for every $j$.
\end{assumption}

With the fixed allocation proportions in Definition \ref{def:design}, Assumptions \ref{ass:maxdev} and \ref{ass:limits} contain the conditions of \citet[Theorem 5]{li2017general}, and Assumption \ref{ass:cov} gives Condition 4 of \citet{zhao2023covariate} for the centered covariates $x_i-\overline x$. The positive limit of $\astar_j$ gives the limits of $\lambda_j$ and $\ahstar$. Under these assumptions, $\astar_j$, $\abarstar$, $\ahstar$, $\lambda_j$ and the finite-population variances of linear combinations of $\Ystar$ with fixed coefficients converge, and Theorem \ref{thm:asy} and Corollary \ref{cor:additive} refer to their limits. The targets $\tau(k)$ and $\ttil_j(k)$ are the finite-population values.

\begin{assumption}\label{ass:alias}
    For every $w\in G\setminus\{\emptyset\}$, $\tau(kw)=0$.
\end{assumption}

By Proposition \ref{prop:bias}, Assumption \ref{ass:alias} is required for every fraction to estimate $\tau(k)$ without bias.

\subsection{Asymptotic results}\label{sec:asyresults}

\begin{theorem}\label{thm:asy}
    Suppose that Assumptions \ref{ass:maxdev} and \ref{ass:limits} hold for the unadjusted version, or Assumption \ref{ass:cov} holds for the adjusted version. Then, the following hold.

    (i) Under $\cD_j$, $N^{1/2}\{\widehat\tau_j-\ttil_j(k)\}\to \mathcal{N}[0,\,4\astar_j-S^2\{\ttil^*_j(k)\}]$ in distribution and $N\vhat_j\to4\astar_j$ in probability. Under $\cD$, $N^{1/2}\{\tauS-\tau(k)\}\to \mathcal{N}[0,\,4\abarstar-S^2\{\tau^*(k)\}]$ in distribution and $N\vhat_{\rm S}\to4\abarstar$ in probability.

    (ii) Under $\cD$, $N\vhat_{\rm W}\to4\ahstar$ in probability. If also Assumption \ref{ass:alias} holds, then under $\cD$, $N^{1/2}\{\tauW-\tau(k)\}\to \mathcal{N}[0,\,4\ahstar-S^2\{\textstyle\sum_{j}\lambda_j\ttil^*_j(k)\}]$ in distribution.

    (iii) $J^{-1}\sum_{j=1}^J4\astar_j=4\abarstar\geq4\ahstar$, with equality if and only if $\astar_1=\cdots=\astar_J$.
\end{theorem}

\S\ref{supp:asy} of the Supplementary Material proves Theorem \ref{thm:asy}. In parts (i) and (ii), $N$ times each variance estimator converges to a limit at least as large as the asymptotic variance. Thus, the Wald intervals, e.g., $\widehat\tau\pm1.96\,\vhat^{1/2}$ at 95\%, are asymptotically conservative for $\ttil_j(k)$ under $\cD_j$ and for $\tau(k)$ under $\cD$. For $\tauW$, this requires Assumption \ref{ass:alias}. Part (iii) compares the limits of $N$ times the estimated variances, which set the widths of the intervals. The true variances contain the terms $-S^2\{\cdot\}$, which the observed data do not identify. Averaged over cosets, $N$ times the estimated variance of a fraction converges to $4\abarstar$, the same limit as for $\tauS$. The weighted estimator uses every coset of the full design and gives less weight to those with larger average cell variance, and $N$ times its estimated variance converges to $4\ahstar$. By \eqref{eq:saturated-formulas} and \eqref{eq:weighted-formulas}, $\vhat_{\rm S}=J^{-2}\sum_j\widehat u_j$, and the inequality between the harmonic and arithmetic means gives $\vhat_{\rm W}\leq\vhat_{\rm S}$ for every assignment. The harmonic mean $\ahstar$ is smaller than $\abarstar$ unless all cosets have the same average cell variance, and the gap is large when one coset is much more variable than another. The weighted estimator requires Assumption \ref{ass:alias}, the assumption under which every fraction estimates $\tau(k)$. It therefore suits experiments in which that assumption is justified before the data are collected, for example by the effect hierarchy principle. Without Assumption \ref{ass:alias}, $\tauS$ still estimates $\tau(k)$, and its interval remains asymptotically conservative.

The next result compares true variances when the aliased effects are the same for every unit.

\begin{assumption}\label{ass:additive}
    For every $w\in G\setminus\{\emptyset\}$ and every unit $i$, $\tau^*_i(kw)=\tau^*(kw)$.
\end{assumption}

\begin{corollary}\label{cor:additive}
    Suppose that the conditions of Theorem \ref{thm:asy} and Assumption \ref{ass:additive} hold. Then,
    \begin{equation}\label{eq:cor}
    \begin{aligned}
        \frac{1}{J}\sum_{j=1}^J\left[4\astar_j-S^2\{\ttil^*_j(k)\}\right] =4\abarstar-S^2\{\tau^*(k)\}\geq4\ahstar-S^2\left\{\sum_{j=1}^J\lambda_j\ttil^*_j(k)\right\},
    \end{aligned}
    \end{equation}
    and the inequality is strict unless $\astar_1=\cdots=\astar_J$. Under Assumption \ref{ass:alias}, the right-hand side is the asymptotic variance of $\tauW$.
\end{corollary}

\S\ref{supp:cor} of the Supplementary Material proves Corollary \ref{cor:additive}. Under Assumption \ref{ass:additive}, each coset contrast of $\Ystar$ differs from $\tau^*_i(k)$ by a constant that is the same for every unit. Thus, every term $S^2\{\cdot\}$ in \eqref{eq:cor} equals $S^2\{\tau^*(k)\}$. The asymptotic variance of a fraction, averaged over cosets, then equals that of $\tauS$. Under Assumption~\ref{ass:alias}, that of $\tauW$ is smaller unless all cosets have the same average cell variance. Without Assumption \ref{ass:additive}, the terms $S^2\{\cdot\}$ in \eqref{eq:cor} can differ, and the order of the true variances can reverse. For example, under Assumption \ref{ass:alias}, if all cosets have the same average cell variance, then $\lambda_j=1/J$ and $\ahstar=\abarstar$. By \eqref{eq:s-average} and \eqref{eq:s-hetavg} of the Supplementary Material applied to $\Ystar$, $\tauW$ then has the same asymptotic variance as $\tauS$, which is larger than the average for the fractions by $\sum_{w\in G\setminus\{\emptyset\}}S^2\{\tau^*(kw)\}$. Without Assumption \ref{ass:alias}, the limit of $\tauW$ in Theorem \ref{thm:asy}(ii) does not apply, and $\tauW$ can be biased for $\tau(k)$ when the cosets have different average cell variances (\S\ref{supp:sim-pretest} of the Supplementary Material).

\subsection{Testing the aliasing assumption}\label{sec:test}

Let $\widehat T=\sum_{j=1}^J(\widehat\delta_j-\tauW)^2/\widehat u_j$, computed under $\cD$, when every $\widehat u_j>0$, and $\widehat T=0$ otherwise. For $0<\gamma<1$, let $\xi_\gamma$ be the upper $\gamma$-quantile of the chi-squared distribution with $J-1$ degrees of freedom. When every $\widehat\psi(z)>0$, $\widehat T$ is the Wald statistic for Assumption \ref{ass:alias} (Lemma \ref{lem:s-gls} of the Supplementary Material).

\begin{proposition}\label{prop:test}
    Suppose that the conditions of Theorem \ref{thm:asy} hold. (i) If Assumption \ref{ass:alias} holds, then $\limsup_{N\to\infty}\bbP(\widehat T>\xi_\gamma)\leq\gamma$, and the limit is $\gamma$ if Assumption \ref{ass:additive} also holds. (ii) If $\tau(kw)$ has a limit for every $w\in G\setminus\{\emptyset\}$ and at least one limit is not zero, then $\bbP(\widehat T>\xi_\gamma)\to1$.
\end{proposition}

\S\ref{supp:test} of the Supplementary Material proves Proposition \ref{prop:test}. Under Assumption \ref{ass:alias}, every $\widehat\delta_j$ estimates $\tau(k)$, and $\widehat T$ measures how far the coset contrasts are from their weighted average. The variation of the target effect across units is common to all cosets and cancels in $\widehat T$. The test is conservative only when the aliased effects of $\Ystar$ vary across units, and it is exact in the limit under Assumption \ref{ass:additive}. A fraction runs one coset and thus does not admit such a test. Notably, Proposition \ref{prop:test} does not cover the pretest estimator, which reports $\tauW$ when the test does not reject and $\tauS$ when it rejects. In fact, intervals after a pretest can under-cover \citep[\S2]{leeb2005model}. Here, when the cosets have unequal average cell variances, a small aliased effect can make $\tauW$ biased and the test often misses it (\S\ref{supp:sim-pretest} of the Supplementary Material). A rejection is evidence against Assumption \ref{ass:alias}, but a non-rejection does not establish it. The test is therefore not a substitute for justifying Assumption \ref{ass:alias} in advance.

\section{Simulation}\label{sec:sim}

Here, we present a simulation that checks Theorems \ref{thm:exact} and \ref{thm:asy}, and \S\ref{supp:sim-table} of the Supplementary Material gives its details. There are $N=400$ units and $K=3$ factors. The generator is $123$, and the target is the main effect of factor 1, which is aliased with the interaction $23$. Scenario 1 has an aliased effect that is zero for every unit and cells with the same variance. Scenario 2 has an aliased effect that varies across units, and Scenario 3 has cosets with different cell variances. Table \ref{tab:sim} reports the theoretical, Monte Carlo, and mean estimated variances, times $N$, and the coverage.

\begin{table}[ht]
    \caption{Simulation results (unadjusted version) for the main effect of factor 1 with $N=400$ over $1{,}000$ replications. The theory column is Theorem \ref{thm:exact}(i), and Theorem \ref{thm:asy}(ii) at the finite population for $\widehat\tau_{\rm W}$. The Monte Carlo standard error of $N\mathrm{var}$ is about $4.5\%$ of the Monte Carlo value, and that of the coverage is at most $0.7$ percentage points.}
    \centering
    \begin{tabular}{ccrrrr}
        \toprule
        Scenario & Estimator & \hd{$N\mathrm{var}$}{(theory)} & \hd{$N\mathrm{var}$}{(Monte Carlo)} & \hd{$N\times$ mean}{$\widehat v$} & \hd{Coverage}{(\%)} \\
        \midrule
        1 & $\widehat\tau_1$ & 4.01 & 3.73 & 5.01 & 97.6 \\
         & $\widehat\tau_2$ & 4.01 & 4.49 & 5.01 & 96.0 \\
         & $\widehat\tau_{\rm S}$ & 4.01 & 3.65 & 5.01 & 98.0 \\
         & $\widehat\tau_{\rm W}$ & 4.01 & 3.70 & 4.99 & 98.3 \\
        2 & $\widehat\tau_1$ & 4.01 & 4.02 & 8.98 & 99.6 \\
         & $\widehat\tau_2$ & 4.01 & 3.80 & 9.06 & 99.8 \\
         & $\widehat\tau_{\rm S}$ & 8.01 & 8.18 & 9.02 & 95.6 \\
         & $\widehat\tau_{\rm W}$ & 8.01 & 8.18 & 8.98 & 95.4 \\
        3 & $\widehat\tau_1$ & 7.86 & 8.23 & 8.86 & 97.1 \\
         & $\widehat\tau_2$ & 1.44 & 1.56 & 2.44 & 97.8 \\
         & $\widehat\tau_{\rm S}$ & 4.65 & 4.89 & 5.65 & 96.8 \\
         & $\widehat\tau_{\rm W}$ & 2.83 & 3.19 & 3.82 & 96.7 \\
    \bottomrule
    \end{tabular}
    \label{tab:sim}
\end{table}

The results match Theorems \ref{thm:exact} and \ref{thm:asy}. In Scenario 2, where the aliased effect varies across units, the fractions have a smaller variance than the full design, as predicted by \eqref{eq:identity}. However, the variance estimators do not detect this gain. In Scenario 3, where the cosets have different cell variances, the mean estimated variances of the two fractions average to that of $\tauS$, as predicted by Theorem \ref{thm:exact}(iii). The weighted estimator has a smaller mean estimated variance than $\tauS$. Further studies in \S\S\ref{supp:sim-figure}--\ref{supp:sim-j4} of the Supplementary Material cover more scenarios, covariate adjustment, a nonzero aliased effect, the test of Proposition \ref{prop:test}, and a design with four cosets.

\section{Discussion}\label{sec:discussion}

We compared regular fractions with the full factorial design under randomization, with the number of units fixed. The precision that a fraction gains on average over cosets comes from the variation of the aliased unit-level effects, which the usual variance estimator does not detect. Under the fraction's assumption, the weighted estimator of the full design has an estimated variance no larger than that of the saturated estimator and, in large samples, no larger than the average for the fractions. The full design can also test the fraction's assumption, but choosing between the two estimators by the test can lose coverage. Nonregular fractions, designs that assign the cosets to fixed groups of units such as successive batches, unequal costs of the treatment combinations, and inference after the pretest are further directions.

\bibliographystyle{jasa3}
\bibliography{fractional_factorial_refs}

\clearpage
\setcounter{section}{0}
\setcounter{equation}{0}
\setcounter{table}{0}
\setcounter{figure}{0}
\setcounter{lemma}{0}
\setcounter{proposition}{0}
\setcounter{corollary}{0}
\renewcommand{\thesection}{S\arabic{section}}
\renewcommand{\theequation}{S\arabic{section}.\arabic{equation}}
\renewcommand{\thetable}{S\arabic{table}}
\renewcommand{\thefigure}{S\arabic{figure}}
\renewcommand{\thelemma}{S\arabic{lemma}}
\renewcommand{\theproposition}{S\arabic{proposition}}
\renewcommand{\thecorollary}{S\arabic{corollary}}
\renewcommand{\theHsection}{S\arabic{section}}
\renewcommand{\theHequation}{S\arabic{section}.\arabic{equation}}
\renewcommand{\theHtable}{S\arabic{table}}
\renewcommand{\theHfigure}{S\arabic{figure}}
\renewcommand{\theHlemma}{S\arabic{lemma}}
\renewcommand{\theHproposition}{S\arabic{proposition}}
\renewcommand{\theHcorollary}{S\arabic{corollary}}

\begin{center}
    {\Large\bfseries Supplementary Material}
\end{center}

\section{Algebra of regular fractions}\label{supp:algebra}

\subsection{Representation over the field with two elements}\label{supp:f2}

This section proves Lemma \ref{lem:alias} of the main text, together with facts about cosets used later. Let $\F$ be the field with two elements. We identify a cell $z\in\cZ$ with the vector $\zeta(z)\in\F^K$ defined by $z_l=(-1)^{\zeta_l(z)}$, and a word $m$ with its indicator vector $\mathsf{e}(m)\in\F^K$. Then,
\begin{equation}\label{eq:s-char}
    g_m(z)=(-1)^{\mathsf{e}(m)^\top\zeta(z)},\qquad \mathsf{e}(mw)=\mathsf{e}(m)+\mathsf{e}(w),
\end{equation}
where the inner product is computed in $\F$. A product of a nonempty subset of the generators equals $\emptyset$ if and only if the corresponding sum of indicator vectors is zero. Thus, the generators of a regular fraction have linearly independent indicator vectors, and the $p\times K$ matrix $M$ with rows $\mathsf{e}(w_1)^\top,\ldots,\mathsf{e}(w_p)^\top$ has rank $p$. The subgroup $G$ corresponds to the row space of $M$. For a sign vector $t$, let $\sigma\in\F^p$ satisfy $t_r=(-1)^{\sigma_r}$. By \eqref{eq:s-char}, the coset defined by $t$ corresponds to the solution set of $M\zeta=\sigma$. Since $M$ has rank $p$, this system is consistent and has $2^{K-p}$ solutions for every $\sigma$. For $w\in G$ with $\mathsf{e}(w)=\sum_{r\in R}\mathsf{e}(w_r)$, every $z$ in the coset $\cC_j$ defined by $t$ satisfies $\mathsf{e}(w_r)^\top\zeta(z)=\sigma_r$ for each $r$, so \eqref{eq:s-char} gives $g_w(z)=\prod_{r\in R}(-1)^{\mathsf{e}(w_r)^\top\zeta(z)}=\prod_{r\in R}(-1)^{\sigma_r}$. Therefore, the value of $g_w$ on $\cC_j$ is
\begin{equation}\label{eq:s-chi}
    \chi_j(w)=\prod_{r\in R}t_r.
\end{equation}
The set $R$ is unique because the rows of $M$ are linearly independent, and $R=\emptyset$ if and only if $w=\emptyset$.

\begin{lemma}\label{lem:s-cosetsum}
    For every word $m$ and every coset $j$,
    \begin{equation*}
        \sum_{z\in\cC_j}g_m(z)=\begin{cases}2^{K-p}\chi_j(m), & m\in G,\\ 0, & m\notin G.\end{cases}
    \end{equation*}
\end{lemma}

\begin{proof}
    If $m\in G$, then $g_m(z)=\chi_j(m)$ for every $z\in\cC_j$, and $\cC_j$ has $2^{K-p}$ elements. If $m\notin G$, then $\mathsf{e}(m)$ is not in the row space of $M$, and the matrix obtained by appending the row $\mathsf{e}(m)^\top$ to $M$ has rank $p+1$. Let $\sigma_j$ be the vector that defines $\cC_j$. For each of the two values in $\F$ prescribed for $\mathsf{e}(m)^\top\zeta$, the system formed by $M\zeta=\sigma_j$ and that equation is consistent and has $2^{K-p-1}$ solutions. By \eqref{eq:s-char},
    \begin{equation*}
        \sum_{z\in\cC_j}g_m(z)=2^{K-p-1}(-1)^0+2^{K-p-1}(-1)^1=0.
    \end{equation*}
\end{proof}

\begin{lemma}\label{lem:s-orth}
    For $w,w'\in G$,
    \begin{equation}\label{eq:s-orth}
    \begin{aligned}
        \frac{1}{J}\sum_{j=1}^J\chi_j(w)=\begin{cases}1, & w=\emptyset,\\ 0, & w\neq\emptyset,\end{cases}\qquad \frac{1}{J}\sum_{j=1}^J\chi_j(w)\chi_j(w')=\begin{cases}1, & w=w',\\ 0, & w\neq w'.\end{cases}
    \end{aligned}
    \end{equation}
\end{lemma}

\begin{proof}
    Let $\mathsf{e}(w)=\sum_{r\in R}\mathsf{e}(w_r)$. As $j$ runs over the $J$ cosets, the sign vector $t$ runs over $\{-1,+1\}^p$. By \eqref{eq:s-chi},
    \begin{equation*}
        \sum_{j=1}^J\chi_j(w)=\sum_{t\in\{-1,+1\}^p}\prod_{r\in R}t_r=2^{p-|R|}\prod_{r\in R}(1-1).
    \end{equation*}
    The right-hand side is $0$ if $R\neq\emptyset$ and $J$ if $R=\emptyset$, and $R=\emptyset$ if and only if $w=\emptyset$. This is the first identity. For the second, $g_w(z)g_{w'}(z)=g_{ww'}(z)$ gives $\chi_j(w)\chi_j(w')=\chi_j(ww')$ with $ww'\in G$. Also, $ww'=\emptyset$ if and only if $w=w'$. The first identity applied to $ww'$ gives the second.
\end{proof}

\begin{lemma}\label{lem:s-union}
    Let $p\geq2$. For two cosets $j\neq j'$, let $H=\{w\in G:\chi_j(w)=\chi_{j'}(w)\}$. Then, $H$ is a subgroup of $G$ with $2^{p-1}$ elements, $k\notin H$, and $\cC_j\cup\cC_{j'}=\{z\in\cZ:g_w(z)=\chi_j(w)\text{ for every }w\in H\}$, which is a coset of the regular fraction with defining contrast subgroup $H$.
\end{lemma}

\begin{proof}
    Let $\phi(w)=\chi_j(w)\chi_{j'}(w)$ for $w\in G$. For $w,w'\in G$, $ww'\in G$, and evaluating $g_w(z)g_{w'}(z)=g_{ww'}(z)$ at any $z\in\cC_j$ gives $\chi_j(ww')=\chi_j(w)\chi_j(w')$. The same holds for $j'$, so $\phi(ww')=\phi(w)\phi(w')$. The two cosets have different sign vectors, so $\phi(w_r)=-1$ for some generator $w_r$. Hence $H=\{w\in G:\phi(w)=1\}$ is closed under products. The map $w\mapsto ww_r$ is one-to-one from $H$ onto $G\setminus H$, so $H$ has $|G|/2=2^{p-1}$ elements. In the representation of \S\ref{supp:f2}, $\mathsf{e}(H)$ is a subspace of the row space of $M$ of dimension $p-1$. Let $M_H$ be a $(p-1)\times K$ matrix whose rows are a basis of $\mathsf{e}(H)$. Its rows are the indicator vectors of $p-1$ generators of $H$, no product of a nonempty subset of which equals $\emptyset$. Since $g_{ww'}=g_wg_{w'}$ and $\chi_j(ww')=\chi_j(w)\chi_j(w')$, a cell $z$ satisfies $g_w(z)=\chi_j(w)$ for every $w\in H$ if and only if it does so for the $p-1$ basis words of $H$. By \eqref{eq:s-char}, these cells correspond to the solutions of $M_H\zeta=\sigma_H$, where the $r$th entry of $\sigma_H\in\F^{p-1}$ satisfies $(-1)^{\sigma_{H,r}}=\chi_j(w)$ for the word $w$ of the $r$th row of $M_H$. Since $M_H$ has rank $p-1$, this system has $2^{K-p+1}$ solutions. Thus, $\mathcal{B}=\{z\in\cZ:g_w(z)=\chi_j(w)\text{ for every }w\in H\}$ has $2^{K-p+1}$ elements. Every $z\in\cC_j$ satisfies $g_w(z)=\chi_j(w)$ for all $w\in G$, so $\cC_j\subseteq\mathcal{B}$. Every $z\in\cC_{j'}$ satisfies $g_w(z)=\chi_{j'}(w)=\chi_j(w)$ for all $w\in H$, so $\cC_{j'}\subseteq\mathcal{B}$. The cosets $\cC_j$ and $\cC_{j'}$ are disjoint with $2^{K-p}$ elements each, so $\mathcal{B}=\cC_j\cup\cC_{j'}$. Finally, $k\notin G$ and $H\subseteq G$ give $k\notin H$.
\end{proof}

\subsection{Proof of Lemma \ref{lem:alias}}\label{supp:lemma1}

For a unit $i$ and a word $m$, let $\theta_i(m)=\sum_{z\in\cZ}g_m(z)Y_i(z)$. The sum below runs over all $2^K$ words $m$, and
\begin{equation*}
    \sum_{m}g_m(z)g_m(z')=\prod_{l=1}^K(1+z_lz'_l)=\begin{cases}2^K, & z=z',\\ 0, & z\neq z'.\end{cases}
\end{equation*}
Multiplying by $Y_i(z')$ and summing over $z'$ gives
\begin{equation}\label{eq:s-inverse}
    Y_i(z)=\frac{1}{2^K}\sum_{m}g_m(z)\,\theta_i(m).
\end{equation}
By \eqref{eq:s-inverse} and $g_k(z)g_m(z)=g_{km}(z)$,
\begin{equation*}
    \sum_{z\in\cC_j}g_k(z)Y_i(z)=\frac{1}{2^K}\sum_{m}\theta_i(m)\sum_{z\in\cC_j}g_{km}(z).
\end{equation*}
By Lemma \ref{lem:s-cosetsum}, the inner sum is $2^{K-p}\chi_j(km)$ when $km\in G$ and zero otherwise. The map $w\mapsto kw$ is a bijection from $G$ onto $\{m:km\in G\}$, because $k(kw)=w$. Hence
\begin{equation*}
    \sum_{z\in\cC_j}g_k(z)Y_i(z)=\frac{1}{2^p}\sum_{w\in G}\chi_j(w)\,\theta_i(kw).
\end{equation*}
Since $k\notin G$, $kw\neq\emptyset$ for every $w\in G$, and \eqref{eq:effect} gives $\theta_i(kw)=2^{K-1}\tau_i(kw)$. Dividing by $2^{K-p-1}$,
\begin{equation*}
    \ttil_{j,i}(k)=\frac{2^{K-1}}{2^{K-p-1}\,2^p}\sum_{w\in G}\chi_j(w)\,\tau_i(kw)=\sum_{w\in G}\chi_j(w)\,\tau_i(kw).
\end{equation*}
This proves Lemma \ref{lem:alias}. Averaging over $j$ and applying the first identity in \eqref{eq:s-orth},
\begin{equation}\label{eq:s-average}
    \frac{1}{J}\sum_{j=1}^J\ttil_{j,i}(k)=\sum_{w\in G}\tau_i(kw)\,\frac{1}{J}\sum_{j=1}^J\chi_j(w)=\tau_i(k).
\end{equation}

\section{Least squares forms of the estimators}\label{supp:ls}

\subsection{Orthogonality of the contrasts in the full design}\label{supp:ols}

Under $\cD$, let $Z_i$ be the cell of unit $i$ and $Y_i$ its observed outcome. For a word $q\neq\emptyset$ with $l\in q$,
\begin{equation}\label{eq:s-fullsum}
    \sum_{z\in\cZ}g_q(z)=\left(\sum_{z_l\in\{-1,1\}}z_l\right)\sum_{z_{-l}}\prod_{l'\in q\setminus\{l\}}z_{l'}=0,
\end{equation}
where $z_{-l}$ runs over the remaining coordinates. Let $m$ and $m'$ be distinct words. Each cell holds $N/2^K$ units, so by \eqref{eq:s-fullsum} with $q=mm'\neq\emptyset$,
\begin{equation*}
    \sum_{i=1}^Ng_m(Z_i)g_{m'}(Z_i)=\frac{N}{2^K}\sum_{z\in\cZ}g_{mm'}(z)=0.
\end{equation*}
Taking $m'=\emptyset$ shows that every contrast column is orthogonal to the intercept. Consider the least squares regression of $Y_i$ on $g_\emptyset(Z_i)=1$ and $g_m(Z_i)$ for $m$ in a set of distinct nonempty words that contains $k$. The regressors are mutually orthogonal, so the coefficient of $g_k(Z_i)$ is
\begin{equation*}
    \frac{\sum_{i}g_k(Z_i)Y_i}{\sum_{i}g_k(Z_i)^2}=\frac{1}{N}\sum_{z\in\cZ}g_k(z)\frac{N}{2^K}\Yhat(z)=\frac{1}{2^K}\sum_{z\in\cZ}g_k(z)\Yhat(z)=\frac{\tauS}{2},
\end{equation*}
where $\Yhat(z)$ is the mean of the observed outcomes in cell $z$.

\subsection{Generalized least squares form of the weighted estimator and the test}\label{supp:gls}

Under $\cD$, let $\widehat\mu=\{\Yhat(z)\}_{z\in\cZ}$ and $\widehat\Psi=\mathrm{diag}\{\widehat\psi(z)\}_{z\in\cZ}$, for either version of Definition \ref{def:est} of the main text. Let $c\in\mathbb{R}^{2^K}$ have entries $c_z=2^{-(K-1)}g_k(z)$, and let $C$ be the $(J-1)\times2^K$ matrix with rows $\{g_{kw}(z)\}_{z\in\cZ}$ for $w\in G\setminus\{\emptyset\}$. For a vector $\mu$ of cell means, $C\mu=0$ states that $\tau(kw)=0$ for every $w\in G\setminus\{\emptyset\}$, which is Assumption \ref{ass:alias} of the main text. Let $\mathcal{X}$ be the $J\times J$ matrix with entries $\mathcal{X}_{jw}=\chi_j(w)$, $w\in G$, and let $\mathcal{X}_0$ be the $J\times(J-1)$ matrix formed by its columns with $w\neq\emptyset$. Let $\widehat\delta=(\widehat\delta_1,\ldots,\widehat\delta_J)^\top$ and $\widehat D=\mathrm{diag}(N\widehat u_1,\ldots,N\widehat u_J)$, as in \S\ref{supp:test}.

\begin{lemma}\label{lem:s-gls}
    Suppose that $\widehat\psi(z)>0$ for every $z\in\cZ$. Let $\widetilde\mu$ minimize $(\widehat\mu-\mu)^\top\widehat\Psi^{-1}(\widehat\mu-\mu)$ over $\mu\in\mathbb{R}^{2^K}$ subject to $C\mu=0$. Then,
    \begin{equation}\label{eq:s-gls}
        c^\top\widetilde\mu=\tauW,\qquad c^\top\left\{\widehat\Psi-\widehat\Psi C^\top(C\widehat\Psi C^\top)^{-1}C\widehat\Psi\right\}c=\vhat_{\rm W}.
    \end{equation}
    With $\widehat\Psi$ replaced by the identity matrix, $c^\top\widetilde\mu=\tauS$. The statistic $\widehat T$ of \S\ref{sec:test} of the main text is the Wald statistic for $C\mu=0$, that is,
    \begin{equation}\label{eq:s-wald}
        \widehat T=(C\widehat\mu)^\top(C\widehat\Psi C^\top)^{-1}C\widehat\mu.
    \end{equation}
\end{lemma}

\begin{proof}
    For $w,w'\in G\setminus\{\emptyset\}$, $g_{kw}(z)g_{kw'}(z)=g_{ww'}(z)$, so \eqref{eq:s-fullsum} gives $\sum_zg_{kw}(z)g_{kw'}(z)=2^K$ if $w=w'$ and $0$ otherwise. Hence $C$ has full row rank. Since $\widehat\Psi$ is positive definite, so is $C\widehat\Psi C^\top$. The objective is strictly convex, and the constraint is linear, so $\widetilde\mu$ is the unique solution of
    \begin{equation*}
        \widehat\Psi^{-1}(\widetilde\mu-\widehat\mu)+C^\top\mathsf{m}=0,\qquad C\widetilde\mu=0,
    \end{equation*}
    for some $\mathsf{m}\in\mathbb{R}^{J-1}$. The first equation gives $\widetilde\mu=\widehat\mu-\widehat\Psi C^\top\mathsf{m}$, and the second then gives $C\widehat\mu=C\widehat\Psi C^\top\mathsf{m}$, so $\mathsf{m}=(C\widehat\Psi C^\top)^{-1}C\widehat\mu$. Hence
    \begin{equation}\label{eq:s-glsfit}
        \begin{gathered}
        \widetilde\mu=\widehat\mu-\widehat\Psi C^\top(C\widehat\Psi C^\top)^{-1}C\widehat\mu,\quad c^\top\widetilde\mu=\ell_*^\top\widehat\mu,\quad \ell_*=c-C^\top\mathsf{r}_*,\quad \mathsf{r}_*=(C\widehat\Psi C^\top)^{-1}C\widehat\Psi c.
        \end{gathered}
    \end{equation}
    Let $\mathcal{L}=\{c+C^\top\mathsf{r}:\mathsf{r}\in\mathbb{R}^{J-1}\}$, so that $\ell_*\in\mathcal{L}$. Write $\mathsf{r}=(\mathsf{r}_w)_{w\in G\setminus\{\emptyset\}}$. For $z\in\cC_j$, $g_{kw}(z)=\chi_j(w)g_k(z)$, so the entry of $\ell=c+C^\top\mathsf{r}$ at $z$ is
    \begin{equation}\label{eq:s-glsentry}
        \ell_z=\frac{g_k(z)}{2^{K-1}}\left\{1+2^{K-1}\sum_{w\in G\setminus\{\emptyset\}}\mathsf{r}_w\chi_j(w)\right\}=\frac{g_k(z)}{2^{K-p-1}}\,\mathsf{b}_j(\mathsf{r}),
    \end{equation}
    where $\mathsf{b}_j(\mathsf{r})=J^{-1}\{1+2^{K-1}\sum_{w\in G\setminus\{\emptyset\}}\mathsf{r}_w\chi_j(w)\}$, because $2^{K-1}=J\,2^{K-p-1}$.
    By \eqref{eq:s-glsentry} and Definition \ref{def:est} of the main text,
    \begin{equation}\label{eq:s-glsclass}
        \ell^\top\widehat\mu=\sum_{j=1}^J\mathsf{b}_j(\mathsf{r})\,\widehat\delta_j,\qquad \ell^\top\widehat\Psi\ell=\sum_{j=1}^J\mathsf{b}_j(\mathsf{r})^2\,\widehat u_j,
    \end{equation}
    and the first identity in \eqref{eq:s-orth} gives $\sum_j\mathsf{b}_j(\mathsf{r})=1$. Conversely, let $\mathsf{b}\in\mathbb{R}^J$ satisfy $\sum_j\mathsf{b}_j=1$, and set $\mathsf{r}_w=2^{-(K-1)}\sum_{j'}\chi_{j'}(w)\mathsf{b}_{j'}$ for $w\in G\setminus\{\emptyset\}$. The second identity in \eqref{eq:s-orth} gives $\mathcal{X}^\top\mathcal{X}=JI_J$. Since $\mathcal{X}$ is square, also $\mathcal{X}\mathcal{X}^\top=JI_J$. The term $w=\emptyset$ of $\sum_{w\in G}\chi_j(w)\sum_{j'}\chi_{j'}(w)\mathsf{b}_{j'}$ is $\sum_{j'}\mathsf{b}_{j'}=1$, so
    \begin{equation*}
        \begin{split}
        1+2^{K-1}\sum_{w\in G\setminus\{\emptyset\}}\mathsf{r}_w\chi_j(w)=\sum_{w\in G}\chi_j(w)\sum_{j'=1}^J\chi_{j'}(w)\mathsf{b}_{j'} =\sum_{j'=1}^J(\mathcal{X}\mathcal{X}^\top)_{jj'}\mathsf{b}_{j'}=J\mathsf{b}_j,
        \end{split}
    \end{equation*}
    that is, $\mathsf{b}(\mathsf{r})=\mathsf{b}$. Therefore, $\mathsf{b}(\mathsf{r})$ ranges over all $\mathsf{b}$ with $\sum_j\mathsf{b}_j=1$ as $\ell$ ranges over $\mathcal{L}$.
    
    Let $\ell=c+C^\top\mathsf{r}\in\mathcal{L}$. Then $\ell=\ell_*+C^\top(\mathsf{r}+\mathsf{r}_*)$. By \eqref{eq:s-glsfit}, $\ell_*^\top\widehat\Psi C^\top=c^\top\widehat\Psi C^\top-c^\top\widehat\Psi C^\top(C\widehat\Psi C^\top)^{-1}C\widehat\Psi C^\top=0$, so
    \begin{equation}\label{eq:s-glsmin}
        \ell^\top\widehat\Psi\ell=\ell_*^\top\widehat\Psi\ell_*+(\mathsf{r}+\mathsf{r}_*)^\top C\widehat\Psi C^\top(\mathsf{r}+\mathsf{r}_*).
    \end{equation}
    Since $C\widehat\Psi C^\top$ is positive definite, \eqref{eq:s-glsmin} shows that $\ell_*$ is the unique minimizer of $\ell^\top\widehat\Psi\ell$ over $\mathcal{L}$. Every $\widehat u_j$ is positive, because every $\widehat\psi(z)$ is. For $\mathsf{b}$ with $\sum_j\mathsf{b}_j=1$, the Cauchy--Schwarz inequality gives
    \begin{equation*}
        1=\left(\sum_{j=1}^J\mathsf{b}_j\widehat u_j^{1/2}\widehat u_j^{-1/2}\right)^2\leq\sum_{j=1}^J\mathsf{b}_j^2\widehat u_j\sum_{j=1}^J\widehat u_j^{-1},
    \end{equation*}
    with equality if and only if $\mathsf{b}_j\propto\widehat u_j^{-1}$, that is, $\mathsf{b}=\widehat\lambda$ of Definition \ref{def:est} of the main text. By \eqref{eq:s-glsclass} and the correspondence between $\mathcal{L}$ and $\mathsf{b}$, the minimizer $\ell_*$ has $\mathsf{b}(\mathsf{r})=\widehat\lambda$, and $\ell_*^\top\widehat\Psi\ell_*=(\sum_j\widehat u_j^{-1})^{-1}=\vhat_{\rm W}$. Then, \eqref{eq:s-glsfit} and \eqref{eq:s-glsclass} give $c^\top\widetilde\mu=\sum_j\widehat\lambda_j\widehat\delta_j=\tauW$. By \eqref{eq:s-glsfit},
    \begin{equation*}
        \ell_*^\top\widehat\Psi\ell_*=c^\top\widehat\Psi c-2c^\top\widehat\Psi C^\top\mathsf{r}_*+\mathsf{r}_*^\top C\widehat\Psi C^\top\mathsf{r}_*=c^\top\widehat\Psi c-c^\top\widehat\Psi C^\top(C\widehat\Psi C^\top)^{-1}C\widehat\Psi c,
    \end{equation*}
    which gives the second identity in \eqref{eq:s-gls}.
    
    With $\widehat\Psi$ replaced by the identity matrix, the same argument applies with $\widehat\psi(z)=1$ for every $z$. Then, every $\widehat u_j$ equals $4^{-(K-p-1)}2^{K-p}$, the minimizing weights are $\mathsf{b}_j=1/J$, and $c^\top\widetilde\mu=J^{-1}\sum_j\widehat\delta_j=\tauS$.
    
    For \eqref{eq:s-wald}, since $g_{kw}(z)=\chi_j(w)g_k(z)$ and $g_k(z)^2=1$ for $z\in\cC_j$, for $w,w'\in G\setminus\{\emptyset\}$,
    \begin{align*}
        (C\widehat\mu)_w&=\sum_{j=1}^J\chi_j(w)\sum_{z\in\cC_j}g_k(z)\Yhat(z)=2^{K-p-1}\sum_{j=1}^J\chi_j(w)\widehat\delta_j,\displaybreak[0]\\
        (C\widehat\Psi C^\top)_{ww'}&=\sum_{j=1}^J\chi_j(w)\chi_j(w')\sum_{z\in\cC_j}\widehat\psi(z)=4^{K-p-1}\sum_{j=1}^J\chi_j(w)\chi_j(w')\widehat u_j.
    \end{align*}
    Therefore, $C\widehat\mu=2^{K-p-1}\mathcal{X}_0^\top\widehat\delta$ and $C\widehat\Psi C^\top=4^{K-p-1}N^{-1}\mathcal{X}_0^\top\widehat D\mathcal{X}_0$, and
    \begin{equation}\label{eq:s-wald1}
        (C\widehat\mu)^\top(C\widehat\Psi C^\top)^{-1}C\widehat\mu=N\widehat\delta^\top\mathcal{X}_0(\mathcal{X}_0^\top\widehat D\mathcal{X}_0)^{-1}\mathcal{X}_0^\top\widehat\delta.
    \end{equation}
    The matrix $\widehat D^{1/2}\mathcal{X}_0(\mathcal{X}_0^\top\widehat D\mathcal{X}_0)^{-1}\mathcal{X}_0^\top\widehat D^{1/2}$ is the orthogonal projection onto the column space of $\widehat D^{1/2}\mathcal{X}_0$. Since $\mathcal{X}^\top\mathcal{X}=JI_J$, the matrix $\mathcal{X}_0$ has rank $J-1$. By the first identity in \eqref{eq:s-orth}, $\mathbf{1}^\top\mathcal{X}_0=0$, so $(\widehat D^{-1/2}\mathbf{1})^\top\widehat D^{1/2}\mathcal{X}_0=0$. The column space of $\widehat D^{1/2}\mathcal{X}_0$ therefore has dimension $J-1$ and is the orthogonal complement of $\widehat D^{-1/2}\mathbf{1}$ in $\mathbb{R}^J$. Hence
    \begin{gather*}
        \widehat D^{1/2}\mathcal{X}_0(\mathcal{X}_0^\top\widehat D\mathcal{X}_0)^{-1}\mathcal{X}_0^\top\widehat D^{1/2}=I_J-\frac{\widehat D^{-1/2}\mathbf{1}\mathbf{1}^\top\widehat D^{-1/2}}{\mathbf{1}^\top\widehat D^{-1}\mathbf{1}},\\
        \mathcal{X}_0(\mathcal{X}_0^\top\widehat D\mathcal{X}_0)^{-1}\mathcal{X}_0^\top =\widehat D^{-1}-\frac{\widehat D^{-1}\mathbf{1}\mathbf{1}^\top\widehat D^{-1}}{\mathbf{1}^\top\widehat D^{-1}\mathbf{1}}.
    \end{gather*}
    Substituting into \eqref{eq:s-wald1}, with $N\widehat D^{-1}=\mathrm{diag}(\widehat u_1^{-1},\ldots,\widehat u_J^{-1})$,
    \begin{equation*}
        (C\widehat\mu)^\top(C\widehat\Psi C^\top)^{-1}C\widehat\mu=\sum_{j=1}^J\frac{\widehat\delta_j^2}{\widehat u_j}-\frac{(\sum_{j=1}^J\widehat\delta_j/\widehat u_j)^2}{\sum_{j=1}^J\widehat u_j^{-1}}=\sum_{j=1}^J\frac{(\widehat\delta_j-\tauW)^2}{\widehat u_j},
    \end{equation*}
    where the last equality expands the square and uses $\tauW=\sum_j(\widehat\delta_j/\widehat u_j)/\sum_j\widehat u_j^{-1}$. The right-hand side is $\widehat T$, because every $\widehat u_j$ is positive.
\end{proof}

For the unadjusted version, $\widehat\psi(z)=s^2(z)/n_z$, and
\begin{equation*}
    \sum_{i=1}^N\frac{\{Y_i-\mu(Z_i)\}^2}{s^2(Z_i)}=\sum_{z\in\cZ}\frac{1}{s^2(z)}\sum_{i:Z_i=z}\{Y_i-\Yhat(z)\}^2+\sum_{z\in\cZ}\frac{\{\Yhat(z)-\mu(z)\}^2}{\widehat\psi(z)}.
\end{equation*}
The first sum does not depend on $\mu$. Thus, $\widetilde\mu$ is the weighted least squares fit of the observed outcome on the cell indicators, with weights $1/s^2(Z_i)$, subject to $C\mu=0$. The $2^K$ contrasts $g_m$ are mutually orthogonal by \eqref{eq:s-fullsum}, so they form a basis of the functions on $\cZ$, and $C\mu=0$ sets to zero the coefficients of $g_{kw}$, $w\in G\setminus\{\emptyset\}$. Hence the restricted fit is the weighted least squares regression on an intercept and the contrasts $g_m(Z_i)$ for every word $m\neq\emptyset$ outside $\{kw:w\in G\setminus\{\emptyset\}\}$. With equal weights, the restricted fit is the least squares fit of \S\ref{supp:ols}, which gives $\tauS$.

\section{Proofs of Proposition \ref{prop:bias} and Theorem \ref{thm:exact}}\label{supp:exact}

\subsection{A variance formula}\label{supp:li-ding}

We use the following result of \citet[Theorem 3]{li2017general}. Suppose that the $N$ units are assigned by complete randomization to a set of cells $\cQ$, with $n_z\geq2$ units in cell $z$ and $\sum_{z\in\cQ}n_z=N$. Let $\Yhat(z)$ be the mean of the observed outcomes in cell $z$, and let $c_z$ be constants. Then,
\begin{equation}\label{eq:s-lidingvar}
\begin{aligned}
    \bbE\left\{\sum_{z\in\cQ}c_z\Yhat(z)\right\}&=\sum_{z\in\cQ}c_z\overline Y(z), \\
    \mathrm{var}\left\{\sum_{z\in\cQ}c_z\Yhat(z)\right\}&=\sum_{z\in\cQ}\frac{c_z^2S^2(z)}{n_z}-\frac{1}{N}S^2\left\{\sum_{z\in\cQ}c_zY(z)\right\},
\end{aligned}
\end{equation}
where $S^2\{\sum_zc_zY(z)\}$ is the finite-population variance of the unit-level values $\sum_zc_zY_i(z)$.

Let $s^2(z)$ be the sample variance of the observed outcomes in cell $z$, and let $A_z$ be the set of units in cell $z$. The units in $A_z$ are a simple random sample of size $n_z$ from the $N$ units. Write
\begin{equation*}
    (n_z-1)\,s^2(z)=\sum_{i\in A_z}\{Y_i(z)-\overline Y(z)\}^2-n_z\{\Yhat(z)-\overline Y(z)\}^2.
\end{equation*}
Each unit is in $A_z$ with probability $n_z/N$, so the expectation of the first term is $n_z(N-1)S^2(z)/N$. By \eqref{eq:s-lidingvar} with $c_z=1$ and all other constants zero, $\mathrm{var}\{\Yhat(z)\}=(1/n_z-1/N)S^2(z)$, so the expectation of the second term is $(1-n_z/N)S^2(z)$. Hence
\begin{equation}\label{eq:s-s2unbiased}
    \bbE\{s^2(z)\}=\frac{1}{n_z-1}\left\{\frac{n_z(N-1)}{N}-\frac{N-n_z}{N}\right\}S^2(z)=S^2(z).
\end{equation}

\subsection{Proof of Proposition \ref{prop:bias}}\label{supp:prop1}

Under $\cD_j$, apply \eqref{eq:s-lidingvar} with $\cQ=\cC_j$ and $c_z=2^{-(K-p-1)}g_k(z)$. Then, $\bbE_j(\widehat\tau_j)=\ttil_j(k)$. Averaging Lemma \ref{lem:alias} over units and separating the term $w=\emptyset$, for which $\chi_j(\emptyset)=1$,
\begin{equation*}
    \bbE_j(\widehat\tau_j)=\tau(k)+\sum_{w\in G\setminus\{\emptyset\}}\chi_j(w)\,\tau(kw).
\end{equation*}
Under $\cD$, apply \eqref{eq:s-lidingvar} with $\cQ=\cZ$ and $c_z=2^{-(K-1)}g_k(z)$. Then, $\bbE(\tauS)=\tau(k)$ by \eqref{eq:effect}.

\subsection{Proof of Theorem \ref{thm:exact}}\label{supp:thm1}

For part (i), under $\cD_j$, each cell of $\cC_j$ holds $n_z=N/2^{K-p}$ units, and $c_z^2=4^{-(K-p-1)}$. The unit-level values of the contrast are $\sum_{z\in\cC_j}c_zY_i(z)=\ttil_{j,i}(k)$. By \eqref{eq:s-lidingvar},
\begin{align*}
    N\mathrm{var}_j(\widehat\tau_j)&=\frac{2^{K-p}}{4^{K-p-1}}\sum_{z\in\cC_j}S^2(z)-S^2\{\ttil_j(k)\}\\
    &=\frac{4}{2^{K-p}}\sum_{z\in\cC_j}S^2(z)-S^2\{\ttil_j(k)\}=4a_j-S^2\{\ttil_j(k)\}.
\end{align*}
Under $\cD$, each cell holds $N/2^K$ units, $c_z^2=4^{-(K-1)}$, and the unit-level values are $\tau_i(k)$. By \eqref{eq:s-lidingvar},
\begin{equation}\label{eq:s-fullvar}
    N\mathrm{var}(\tauS)=\frac{4}{2^K}\sum_{z\in\cZ}S^2(z)-S^2\{\tau(k)\}=4\abar-S^2\{\tau(k)\}.
\end{equation}
The last equality holds because the $J$ cosets partition $\cZ$ and each has $2^{K-p}$ cells, so that $2^{-K}\sum_{z\in\cZ}S^2(z)=J^{-1}\sum_{j}a_j$.

For part (ii), let $\Delta_i(m)=\tau_i(m)-\tau(m)$. Subtracting the average over units of Lemma \ref{lem:alias} from Lemma \ref{lem:alias} gives
\begin{equation*}
    \ttil_{j,i}(k)-\ttil_j(k)=\sum_{w\in G}\chi_j(w)\,\Delta_i(kw).
\end{equation*}
Squaring, summing over units and averaging over cosets,
\begin{equation*}
    \frac{1}{J}\sum_{j=1}^JS^2\{\ttil_j(k)\}=\frac{1}{N-1}\sum_{i=1}^N\sum_{w,w'\in G}\Delta_i(kw)\,\Delta_i(kw')\,\frac{1}{J}\sum_{j=1}^J\chi_j(w)\chi_j(w').
\end{equation*}
By the second identity in \eqref{eq:s-orth}, only the terms with $w=w'$ remain, and
\begin{equation}\label{eq:s-hetavg}
    \frac{1}{J}\sum_{j=1}^JS^2\{\ttil_j(k)\}=\frac{1}{N-1}\sum_{i=1}^N\sum_{w\in G}\Delta_i(kw)^2=\sum_{w\in G}S^2\{\tau(kw)\}.
\end{equation}
Averaging part (i) over $j$ and using \eqref{eq:s-hetavg},
\begin{equation*}
    \frac{1}{J}\sum_{j=1}^JN\mathrm{var}_j(\widehat\tau_j)=4\abar-S^2\{\tau(k)\}-\sum_{w\in G\setminus\{\emptyset\}}S^2\{\tau(kw)\}.
\end{equation*}
By \eqref{eq:s-fullvar}, the first two terms on the right are $N\mathrm{var}(\tauS)$. This is \eqref{eq:identity}.

For part (iii), $\widehat\psi(z)=s^2(z)/n_z$, so \eqref{eq:s-s2unbiased} gives
\begin{gather*}
    N\bbE_j(\vhat_j)=\frac{N}{4^{K-p-1}}\sum_{z\in\cC_j}\frac{S^2(z)}{N/2^{K-p}}=4a_j,\qquad N\bbE(\vhat_{\rm S})=\frac{N}{4^{K-1}}\sum_{z\in\cZ}\frac{S^2(z)}{N/2^{K}}=4\abar.
\end{gather*}
Averaging the first identity over $j$ gives $J^{-1}\sum_{j}N\bbE_j(\vhat_j)=4\abar=N\bbE(\vhat_{\rm S})$.

For part (iv), conditional on $U=j$, the estimator $\widehat\tau_U$ is $\widehat\tau_j$ under $\cD_j$. By Proposition \ref{prop:bias} and Lemma \ref{lem:alias} averaged over units, its conditional mean is $\ttil_j(k)=\tau(k)+\sum_{w\neq\emptyset}\chi_j(w)\tau(kw)$, where here and below $w$ runs over $G\setminus\{\emptyset\}$. The first identity in \eqref{eq:s-orth} gives
\begin{equation}\label{eq:s-randmean}
    \bbE_U(\widehat\tau_U)=\frac{1}{J}\sum_{j=1}^J\ttil_j(k)=\tau(k)+\sum_{w\neq\emptyset}\tau(kw)\,\frac{1}{J}\sum_{j=1}^J\chi_j(w)=\tau(k).
\end{equation}
By the law of total variance and \eqref{eq:s-randmean},
\begin{equation*}
    \mathrm{var}_U(\widehat\tau_U)=\frac{1}{J}\sum_{j=1}^J\mathrm{var}_j(\widehat\tau_j)+\frac{1}{J}\sum_{j=1}^J\left\{\sum_{w\neq\emptyset}\chi_j(w)\tau(kw)\right\}^2.
\end{equation*}
By the second identity in \eqref{eq:s-orth}, the last term is
\begin{equation*}
    \sum_{w\neq\emptyset}\sum_{w'\neq\emptyset}\tau(kw)\tau(kw')\,\frac{1}{J}\sum_{j=1}^J\chi_j(w)\chi_j(w')=\sum_{w\neq\emptyset}\tau(kw)^2.
\end{equation*}
Multiplying by $N$ gives part (iv).

\section{Proof of Theorem \ref{thm:asy}}\label{supp:asy}

\subsection{Two limit results}\label{supp:limits}

Consider either design, with cell set $\cQ$ equal to $\cC_j$ or $\cZ$, and let $e_z=n_z/N$. Under Definition \ref{def:design} of the main text, $e_z$ equals $2^{-(K-p)}$ or $2^{-K}$. The results of this subsection hold for any allocation proportions $e_z$ that are fixed and positive, which is used in \S\ref{supp:unequal-asy}. For $z,z'\in\cQ$, let $S(z,z')$ and $S_*(z,z')$ be the finite-population covariances of $Y(z)$ and $Y(z')$ and of $\Ystar(z)$ and $\Ystar(z')$, and write $S^2_*(z)=S_*(z,z)$, which is $S^2\{\Ystar(z)\}$. For the unadjusted version, $S_*(z,z')=S(z,z')$, which has a finite limit by Assumption \ref{ass:limits}. For the adjusted version, $\Ystar_i(z)=Y_i(z)-\widetilde x_i^\top\beta(z)$ with $\widetilde x_i=x_i-\overline x$, and $S_{xY(z)}=S_{xx}\beta(z)$, so
\begin{equation}\label{eq:s-covstar}
\begin{split}
    S_*(z,z')&=S(z,z')-\beta(z')^\top S_{xY(z)}-\beta(z)^\top S_{xY(z')}+\beta(z)^\top S_{xx}\beta(z')\\
    &=S(z,z')-S_{xY(z)}^\top S_{xx}^{-1}S_{xY(z')}.
\end{split}
\end{equation}
By Assumption \ref{ass:cov}(i), $S(z,z')$, $S_{xY(z)}$, $S_{xY(z')}$ and $S_{xx}$ have finite limits, and by Assumption \ref{ass:cov}(ii) the limit of $S_{xx}$ is nonsingular. Matrix inversion is continuous at a nonsingular matrix, so \eqref{eq:s-covstar} has a finite limit. In both versions, for coefficients $c_z$ that converge to finite limits,
\begin{equation*}
    \astar_j=\frac{1}{2^{K-p}}\sum_{z\in\cC_j}S_*(z,z),\qquad S^2\left\{\sum_{z\in\cQ}c_z\Ystar(z)\right\}=\sum_{z\in\cQ}\sum_{z'\in\cQ}c_zc_{z'}S_*(z,z')
\end{equation*}
converge to finite limits. The limit of every $\astar_j$ is positive by Assumption \ref{ass:limits} or \ref{ass:cov}(iv). Therefore, $\abarstar$, $\ahstar=J/\sum_j(\astar_j)^{-1}$ and $\lambda_j=(\astar_j)^{-1}/\sum_{j'}(\astar_{j'})^{-1}$ converge, being continuous functions of $(\astar_1,\ldots,\astar_J)$ at its limit. As in \S\ref{sec:assumptions} of the main text, these variance quantities stand for their limits below, while $\overline Y(z)$, $\tau(k)$ and $\ttil_j(k)$ are finite-population values. Under the conditions of Theorem \ref{thm:asy}, the following hold for every fixed vector of constants $(c_z)_{z\in\cQ}$:
\begin{equation}\label{eq:s-clt}
\begin{aligned}    
    N^{1/2}\sum_{z\in\cQ}c_z\{\Yhat(z)-\overline Y(z)\} \to\mathcal{N}\left[0,\ \sum_{z\in\cQ}\frac{c_z^2S^2_*(z)}{e_z}-S^2\left\{\sum_{z\in\cQ}c_z\Ystar(z)\right\}\right]
\end{aligned}
\end{equation}
in distribution, and
\begin{equation}\label{eq:s-psi}
    N\widehat\psi(z)-\frac{S^2_*(z)}{e_z}\to0\quad\text{in probability},\qquad z\in\cQ.
\end{equation}

For the unadjusted version, \eqref{eq:s-clt} follows from \citet[Theorem 5]{li2017general}, applied to the vector of cell means. Its conditions are Assumptions \ref{ass:maxdev} and \ref{ass:limits} and the allocation proportions $e_z$, which are fixed and positive. Proposition 3 of \citet{li2017general} gives $s^2(z)-S^2(z)\to0$ in probability, and $N\widehat\psi(z)=s^2(z)/e_z$, which gives \eqref{eq:s-psi}.

For the adjusted version, the cell estimates $\Yhat(z)$ are the intercepts of the fully interacted least squares fit of \citet{zhao2023covariate}, with covariates centered at $\overline x$. Each row of the regressor matrix of that fit is nonzero only in the block of the cell of its unit. Thus, the cross-product matrix and the middle matrix of its Eicker--Huber--White (EHW) covariance estimate, $\widehat\Psi$, are both block diagonal across cells, and $\widehat\Psi$ is diagonal with entries $\widehat\psi(z)$. \citet{zhao2023covariate} center the covariates, so we apply their results to $\widetilde x_i=x_i-\overline x$. Their Condition 4 holds for $\widetilde x$, because $S_{\widetilde x\widetilde x}=S_{xx}$, the first two moments of $\widetilde x_i$ and of $\widetilde x_iY_i(z)=x_iY_i(z)-\overline xY_i(z)$ have limits by Assumption \ref{ass:cov}(i), and
\begin{gather*}
    \frac{1}{N}\sum_{i=1}^N\|\widetilde x_iY_i(z)\|^4\leq\frac{8}{N}\sum_{i=1}^N\|x_iY_i(z)\|^4+\frac{8\|\overline x\|^4}{N}\sum_{i=1}^NY_i(z)^4,\\ \frac{1}{N}\sum_{i=1}^N\|\widetilde x_i\|^4\leq\frac{8}{N}\sum_{i=1}^N\|x_i\|^4+8\|\overline x\|^4,
\end{gather*}
which are bounded by Assumption \ref{ass:cov}(i) and (iii). Lemma 1(i) of \citet{zhao2023covariate} then gives $N^{1/2}\{\Yhat-\overline Y\}\to\mathcal{N}(0,V)$ in distribution with $V=\mathrm{diag}\{S^2_*(z)/e_z\}-S_*$. Here, $S_*$ is the finite-population covariance matrix of the residuals $\{Y_i(z)-\beta(z)^\top\widetilde x_i:z\in\cQ\}$, and their regression coefficient for cell $z$ is $\beta(z)$. These residuals are $\Ystar_i(z)$, so $c^\top S_*c=S^2\{\sum_zc_z\Ystar(z)\}$. This gives \eqref{eq:s-clt}. The same lemma gives $N\widehat\Psi-V=S_*+o_P(1)$, so $N\widehat\Psi=\mathrm{diag}\{S^2_*(z)/e_z\}+o_P(1)$, which gives \eqref{eq:s-psi}. Under Assumption \ref{ass:cov}(iii),
\begin{align*}
    \max_{1\leq i\leq N}\frac{\|x_i-\overline x\|^2}{N}&\leq\frac{2}{N}\max_{1\leq i\leq N}\|x_i\|^2+\frac{2\|\overline x\|^2}{N}\displaybreak[0]\\
    &\leq\frac{2}{N}\left(\sum_{i=1}^N\|x_i\|^4\right)^{1/2}+\frac{2\|\overline x\|^2}{N}=\frac{O(N^{1/2})+O(1)}{N}\to0,
\end{align*}
where $\overline x$ is bounded by Assumption \ref{ass:cov}(i). The finite-population covariance matrix of $x$ has a limit by Assumption \ref{ass:cov}(i), and $e_z$ is fixed and positive. Thus, $x$ satisfies the conditions of \citet[Theorem 5]{li2017general}, and their Proposition 3 shows that the within-cell sample covariance matrix $\widehat S_{xx}(z)$ of $x$ in cell $z$ converges in probability to the limit $S_{xx,\infty}$ of $S_{xx}$, which is nonsingular. The fit in cell $z$ has full rank if and only if $\widehat S_{xx}(z)$ is nonsingular. Let $\mathcal{E}_N$ be the event that every within-cell fit has full rank, and let $\Xi_0>0$ be the smallest eigenvalue of $S_{xx,\infty}$. By Weyl's inequality, the smallest eigenvalue of $\widehat S_{xx}(z)$ differs from $\Xi_0$ by at most the spectral norm $\|\widehat S_{xx}(z)-S_{xx,\infty}\|$. Hence
\begin{align*}
    \bbP(\mathcal{E}_N^{\rm c})&\leq\sum_{z\in\cQ}\bbP\{\text{the smallest eigenvalue of }\widehat S_{xx}(z)\text{ is at most }\Xi_0/2\}\displaybreak[0]\\
    &\leq\sum_{z\in\cQ}\bbP\left(\|\widehat S_{xx}(z)-S_{xx,\infty}\|\geq\Xi_0/2\right)\to0.
\end{align*}
Off $\mathcal{E}_N$, Definition \ref{def:est} uses the unadjusted estimates in the cells whose fits are rank-deficient. For random variables $\Upsilon_N$ and $\Upsilon'_N$ that are equal on $\mathcal{E}_N$, and every $\mathsf{t}>0$,
\begin{equation}\label{eq:s-event}
    \bbP(|\Upsilon_N-\Upsilon'_N|>\mathsf{t})\leq\bbP(\mathcal{E}_N^{\rm c})\to0,
\end{equation}
so $\Upsilon_N-\Upsilon'_N=o_P(1)$. The same bound holds with $\mathcal{E}_N$ replaced by any event whose probability tends to one. Hence \eqref{eq:s-clt} and \eqref{eq:s-psi} hold for the estimators of Definition \ref{def:est} if they hold for the fitted ones.

\subsection{Proof of parts (i)--(iii)}\label{supp:thm2}

For part (i) under $\cD_j$, take $\cQ=\cC_j$ and $c_z=2^{-(K-p-1)}g_k(z)$. Then, $\sum_zc_z\overline Y(z)=\ttil_j(k)$ and $\sum_zc_z\Ystar_i(z)=\ttil^*_{j,i}(k)$, and
\begin{equation}\label{eq:s-varj}
    \sum_{z\in\cC_j}\frac{c_z^2S^2_*(z)}{e_z}=\frac{2^{K-p}}{4^{K-p-1}}\sum_{z\in\cC_j}S^2_*(z)=4\astar_j.
\end{equation}
Therefore, \eqref{eq:s-clt} gives the stated limit for $\widehat\tau_j$. By \eqref{eq:s-psi} and \eqref{eq:s-varj},
\begin{equation*}
    N\vhat_j=\frac{1}{4^{K-p-1}}\sum_{z\in\cC_j}N\widehat\psi(z)=\frac{1}{4^{K-p-1}}\sum_{z\in\cC_j}\frac{S^2_*(z)}{e_z}+o_P(1)=4\astar_j+o_P(1),
\end{equation*}
since the sum has $2^{K-p}$ terms. Under $\cD$, take $\cQ=\cZ$, $c_z=2^{-(K-1)}g_k(z)$ and $e_z=2^{-K}$. Then, $\sum_zc_z\overline Y(z)=\tau(k)$, $\sum_zc_z\Ystar_i(z)=\tau^*_i(k)$, and $\sum_{z\in\cZ}S^2_*(z)=2^{K-p}\sum_j\astar_j=2^K\abarstar$, so
\begin{gather*}
    \sum_{z\in\cZ}\frac{c_z^2S^2_*(z)}{e_z}=\frac{2^K}{4^{K-1}}\sum_{z\in\cZ}S^2_*(z)=4\abarstar,\qquad N\vhat_{\rm S}=\frac{1}{4^{K-1}}\sum_{z\in\cZ}N\widehat\psi(z)=4\abarstar+o_P(1),
\end{gather*}
where the second identity uses \eqref{eq:s-psi}. Thus, \eqref{eq:s-clt} gives the stated limit for $\tauS$.

For part (ii), work under $\cD$, so $\cQ=\cZ$ and $e_z=2^{-K}$. For each $j$, \eqref{eq:s-clt} with $c^{(j)}_z=2^{-(K-p-1)}g_k(z)1(z\in\cC_j)$ gives convergence in distribution of $N^{1/2}\{\widehat\delta_j-\ttil_j(k)\}$. A sequence that converges in distribution is tight, so
\begin{equation}\label{eq:s-deltaclt}
    N^{1/2}\{\widehat\delta_j-\ttil_j(k)\}=O_P(1),\qquad j=1,\ldots,J.
\end{equation}
By \eqref{eq:s-psi},
\begin{equation}\label{eq:s-uhat}
    N\widehat u_j=\frac{1}{4^{K-p-1}}\sum_{z\in\cC_j}N\widehat\psi(z)=\frac{2^K}{4^{K-p-1}}\sum_{z\in\cC_j}S^2_*(z)+o_P(1)=4J\,\astar_j+o_P(1),
\end{equation}
because $2^K\,2^{K-p}/4^{K-p-1}=4\cdot2^p$. The limits of $\astar_1,\ldots,\astar_J$ are positive by Assumption \ref{ass:limits} or \ref{ass:cov}. Therefore, by \eqref{eq:s-uhat},
\begin{equation}\label{eq:s-upos4}
    \bbP\left(\min_{1\leq j\leq J}\widehat u_j\leq0\right)\leq\sum_{j=1}^J\bbP\left(|N\widehat u_j-4J\astar_j|\geq4J\astar_j\right)\to0,
\end{equation}
and, on the event that every $\widehat u_j$ is positive, $\widehat\lambda_j=(N\widehat u_j)^{-1}/\sum_{j'}(N\widehat u_{j'})^{-1}$ and $N\vhat_{\rm W}=\{\sum_{j}(N\widehat u_{j})^{-1}\}^{-1}$. Outside this event, Definition \ref{def:est} sets $\widehat\lambda_j=1/J$. By \eqref{eq:s-event} with $\mathcal{E}_N$ replaced by this event, these values do not affect limits in probability. The continuous mapping theorem and \eqref{eq:s-uhat} give
\begin{equation}\label{eq:s-lambda}
\begin{aligned}
    \widehat\lambda_j-\lambda_j&=o_P(1),\\ 
    N\vhat_{\rm W}&=\left\{\sum_{j=1}^J(N\widehat u_j)^{-1}\right\}^{-1}=\left\{\sum_{j=1}^J(4J\astar_j)^{-1}\right\}^{-1}+o_P(1)=4\ahstar+o_P(1).
\end{aligned}
\end{equation}
This gives the limit of $N\vhat_{\rm W}$. For the limit of $\tauW$, suppose that Assumption \ref{ass:alias} holds. Then, Lemma \ref{lem:alias} averaged over units gives $\ttil_j(k)=\tau(k)$ for every $j$. Since $\sum_j\widehat\lambda_j=\sum_j\lambda_j=1$,
\begin{equation}\label{eq:s-decomp}
\begin{aligned}
    N^{1/2}\{\tauW-\tau(k)\} =\sum_{j=1}^J\lambda_jN^{1/2}\{\widehat\delta_j-\tau(k)\}+\sum_{j=1}^J(\widehat\lambda_j-\lambda_j)\,N^{1/2}\{\widehat\delta_j-\tau(k)\}.
\end{aligned}
\end{equation}
By \eqref{eq:s-deltaclt} and \eqref{eq:s-lambda}, each term of the second sum is $o_P(1)\,O_P(1)=o_P(1)$. The first sum is the left-hand side of \eqref{eq:s-clt} for $c_z=\lambda_{j(z)}2^{-(K-p-1)}g_k(z)$, where $j(z)$ is the coset containing $z$. These $c_z$ are fixed, because $\lambda_j$ stands for its limit. For these constants, $\sum_zc_z\Ystar_i(z)=\sum_j\lambda_j\ttil^*_{j,i}(k)$ and
\begin{equation*}
    \begin{split}
    \sum_{z\in\cZ}\frac{c_z^2S^2_*(z)}{e_z}&=\sum_{j=1}^J\lambda_j^2\,\frac{2^K}{4^{K-p-1}}\sum_{z\in\cC_j}S^2_*(z)=4J\sum_{j=1}^J\lambda_j^2\astar_j=\frac{4J}{\sum_{j'=1}^J(\astar_{j'})^{-1}}=4\ahstar,
    \end{split}
\end{equation*}
where the third equality uses $\lambda_j=(\astar_j)^{-1}/\sum_{j'}(\astar_{j'})^{-1}$. Slutsky's theorem applied to \eqref{eq:s-decomp} gives the stated limit for $\tauW$.

For part (iii), the Cauchy--Schwarz inequality gives
\begin{equation*}
    J^2=\left\{\sum_{j=1}^J(\astar_j)^{1/2}(\astar_j)^{-1/2}\right\}^2\leq\sum_{j=1}^J\astar_j\sum_{j=1}^J(\astar_j)^{-1},
\end{equation*}
which rearranges to $\abarstar\geq\ahstar$. Equality holds if and only if $(\astar_j)^{1/2}$ is proportional to $(\astar_j)^{-1/2}$, that is, if and only if $\astar_1=\cdots=\astar_J$.

\section{Proof of Corollary \ref{cor:additive}}\label{supp:cor}

Lemma \ref{lem:alias} holds for any potential outcomes, so it holds for $\Ystar$. Under Assumption \ref{ass:additive}, every term with $w\neq\emptyset$ is the same for all units, and $\chi_j(\emptyset)=1$. Subtracting the average over units from Lemma \ref{lem:alias} for $\Ystar$,
\begin{equation*}
    \ttil^*_{j,i}(k)-\ttil^*_j(k)=\tau^*_i(k)-\tau^*(k),\qquad j=1,\ldots,J.
\end{equation*}
Hence $S^2\{\ttil^*_j(k)\}=S^2\{\tau^*(k)\}$ for every $j$. Since $\sum_j\lambda_j=1$, also
\begin{equation*}
    \sum_{j=1}^J\lambda_j\ttil^*_{j,i}(k)-\sum_{j=1}^J\lambda_j\ttil^*_j(k)=\tau^*_i(k)-\tau^*(k),
\end{equation*}
so $S^2\{\sum_j\lambda_j\ttil^*_j(k)\}=S^2\{\tau^*(k)\}$. Substituting into \eqref{eq:cor}, its left-hand side is $4\abarstar-S^2\{\tau^*(k)\}$ and its right-hand side is $4\ahstar-S^2\{\tau^*(k)\}$. Theorem \ref{thm:asy}(iii) gives the inequality and the condition for equality. Under Assumption \ref{ass:alias}, Theorem \ref{thm:asy}(ii) shows that the right-hand side is the asymptotic variance of $\tauW$.

\section{Proof of Proposition \ref{prop:test}}\label{supp:test}

Work under $\cD$, so $\cQ=\cZ$ and $e_z=2^{-K}$. Let $\mathbf{1}$ be the vector of $J$ ones, and let $\widehat\delta=(\widehat\delta_1,\ldots,\widehat\delta_J)^\top$, $\widehat\lambda=(\widehat\lambda_1,\ldots,\widehat\lambda_J)^\top$ and $\lambda=(\lambda_1,\ldots,\lambda_J)^\top$. For $j=1,\ldots,J$, let $\pi_j=\ttil_j(k)-\tau(k)$, so that $\pi_j=\sum_{w\in G\setminus\{\emptyset\}}\chi_j(w)\tau(kw)$ by Lemma \ref{lem:alias}. Let $\pi=(\pi_1,\ldots,\pi_J)^\top$ and $X=N^{1/2}\{\widehat\delta-\tau(k)\mathbf{1}-\pi\}$. For a fixed $\mathsf{s}\in\mathbb{R}^J$, the linear combination $\mathsf{s}^\top X$ is the left-hand side of \eqref{eq:s-clt} with $c_z=\mathsf{s}_{j(z)}2^{-(K-p-1)}g_k(z)$. For these constants, as in the proof of Theorem \ref{thm:asy}(ii),
\begin{equation*}
    \sum_{z\in\cZ}\frac{c_z^2S^2_*(z)}{e_z}=4J\sum_{j=1}^J\mathsf{s}_j^2\astar_j,\qquad S^2\left\{\sum_{z\in\cZ}c_z\Ystar(z)\right\}=\mathsf{s}^\top\Sigma_*\mathsf{s},
\end{equation*}
where $\Sigma_*$ is the finite-population covariance matrix of $(\ttil^*_{1,i}(k),\ldots,\ttil^*_{J,i}(k))$. By the Cram\'er--Wold device,
\begin{equation}\label{eq:s-vecclt}
    X\to\mathcal{N}(0,\,D-\Sigma_*)\quad\text{in distribution},\quad D=\mathrm{diag}(4J\astar_1,\ldots,4J\astar_J).
\end{equation}
Let $\mathcal{F}_N$ be the event that every $\widehat u_j$ is positive. By \eqref{eq:s-upos4}, $\bbP(\mathcal{F}_N)\to1$. Off $\mathcal{F}_N$, $\widehat T=0$ by its definition. Let $\widehat D=\mathrm{diag}(N\widehat u_1,\ldots,N\widehat u_J)$ on $\mathcal{F}_N$ and $\widehat D=D$ off it. On $\mathcal{F}_N$, $\widehat\lambda=\widehat D^{-1}\mathbf{1}/(\mathbf{1}^\top\widehat D^{-1}\mathbf{1})$. By \eqref{eq:s-uhat}, which does not use Assumption \ref{ass:alias}, $\widehat D-D=o_P(1)$, and $D$ is nonsingular. Let $\widehat T'$ be the right-hand side of \eqref{eq:s-tshift} below, computed with this $\widehat D$ and with $\widehat\lambda$ replaced by $\widehat D^{-1}\mathbf{1}/(\mathbf{1}^\top\widehat D^{-1}\mathbf{1})$, which equals $\widehat\lambda$ on $\mathcal{F}_N$ and $\lambda$ off it. Below, $\widehat\lambda$ denotes this vector. On $\mathcal{F}_N$, $\widehat T'=\widehat T$, because $\sum_j\widehat\lambda_j=1$, so adding the same constant to every $\widehat\delta_j$ adds it to $\tauW$, and $\widehat T$ does not change. By \eqref{eq:s-event} with $\mathcal{E}_N$ replaced by $\mathcal{F}_N$, $\widehat T-\widehat T'=o_P(1)$, so $\widehat T$ and $\widehat T'$ have the same limits in distribution and in probability by Slutsky's theorem, and we write $\widehat T$ for $\widehat T'$ below. Then,
\begin{equation}\label{eq:s-tshift}
    \widehat T=\sum_{j=1}^J\frac{\{\widehat\delta_j-\tau(k)-\sum_{j'}\widehat\lambda_{j'}(\widehat\delta_{j'}-\tau(k))\}^2}{\widehat u_j}.
\end{equation}

For part (i), Assumption \ref{ass:alias} gives $\pi=0$, so $X=N^{1/2}\{\widehat\delta-\tau(k)\mathbf{1}\}$. The vector with entries $N^{1/2}\{\widehat\delta_j-\tau(k)-\sum_{j'}\widehat\lambda_{j'}(\widehat\delta_{j'}-\tau(k))\}$ is $(I-\mathbf{1}\widehat\lambda^\top)X$, and \eqref{eq:s-tshift} gives
\begin{equation}\label{eq:s-quadform}
\begin{aligned}
    \widehat T=X^\top(I-\widehat\lambda\mathbf{1}^\top)\widehat D^{-1}(I-\mathbf{1}\widehat\lambda^\top)X=X^\top\widehat PX,\qquad\widehat P=\widehat D^{-1}-\frac{\widehat D^{-1}\mathbf{1}\mathbf{1}^\top\widehat D^{-1}}{\mathbf{1}^\top\widehat D^{-1}\mathbf{1}},
\end{aligned}
\end{equation}
where the second equality uses $\widehat D^{-1}\mathbf{1}=(\mathbf{1}^\top\widehat D^{-1}\mathbf{1})\,\widehat\lambda$. Let $P$ be $\widehat P$ with $\widehat D$ replaced by $D$. Since $\widehat D-D=o_P(1)$ and $D$ is nonsingular, $\widehat P-P=o_P(1)$. By \eqref{eq:s-vecclt}, Slutsky's theorem and the continuous mapping theorem applied to \eqref{eq:s-quadform}, $\widehat T\to X_\infty^\top PX_\infty$ in distribution, where $X_\infty\sim\mathcal{N}(0,D-\Sigma_*)$. Let $h=D^{-1/2}\mathbf{1}$ and $\Pi =I-hh^\top/(h^\top h)$, which is symmetric and idempotent with rank $J-1$. Write $\Pi =LL^\top$ with $L$ a $J\times(J-1)$ matrix with orthonormal columns. With $\Phi=L^\top D^{-1/2}X_\infty$,
\begin{equation}\label{eq:s-zq}
    X_\infty^\top PX_\infty=X_\infty^\top D^{-1/2}\Pi D^{-1/2}X_\infty=\|\Phi\|^2,\qquad \Phi\sim\mathcal{N}(0,\Omega),
\end{equation}
where $\Omega=I_{J-1}-L^\top D^{-1/2}\Sigma_*D^{-1/2}L=L^\top D^{-1/2}(D-\Sigma_*)D^{-1/2}L$. The second form shows that $\Omega$ is positive semidefinite, because $D-\Sigma_*$ is a limiting covariance matrix by \eqref{eq:s-vecclt}.
Since $\Sigma_*$ is positive semidefinite, $I_{J-1}-\Omega$ is positive semidefinite. Let $\Gamma\sim\mathcal{N}(0,I_{J-1})$. Then, $\Omega^{1/2}\Gamma$ has the distribution of $\Phi$, and
\begin{equation}\label{eq:s-coupling}
    \|\Omega^{1/2}\Gamma\|^2=\Gamma^\top\Omega\Gamma\leq\Gamma^\top\Gamma.
\end{equation}
Here, $\Gamma^\top\Gamma$ has the chi-squared distribution with $J-1$ degrees of freedom, which has no atoms. The set $[\xi_\gamma,\infty)$ is closed, so the portmanteau theorem, \eqref{eq:s-zq} and \eqref{eq:s-coupling} give
\begin{equation*}
    \begin{split}
    \limsup_{N\to\infty}\bbP(\widehat T>\xi_\gamma)&\leq\limsup_{N\to\infty}\bbP(\widehat T\geq\xi_\gamma) \leq\bbP(\Gamma^\top\Omega\Gamma\geq\xi_\gamma)\leq\bbP(\Gamma^\top\Gamma\geq\xi_\gamma)=\gamma.
    \end{split}
\end{equation*}
If Assumption \ref{ass:additive} also holds, the proof of Corollary \ref{cor:additive} gives $\ttil^*_{j,i}(k)-\ttil^*_j(k)=\tau^*_i(k)-\tau^*(k)$ for every $j$. Thus, $\Sigma_*=S^2\{\tau^*(k)\}\,\mathbf{1}\mathbf{1}^\top$ and $D^{-1/2}\Sigma_*D^{-1/2}=S^2\{\tau^*(k)\}\,hh^\top$. Since $L^\top h=0$, $\Omega=I_{J-1}$, and $X_\infty^\top PX_\infty$ has the chi-squared distribution with $J-1$ degrees of freedom. This distribution is continuous, so $\bbP(\widehat T>\xi_\gamma)\to\gamma$.

For part (ii), $\pi$ converges to a limit $\pi_\infty$, because each $\tau(kw)$ with $w\neq\emptyset$ converges. For $w'\in G\setminus\{\emptyset\}$, the identities in \eqref{eq:s-orth} give
\begin{equation}\label{eq:s-invert}
    \frac{1}{J}\sum_{j=1}^J\chi_j(w')\,\pi_j=\sum_{w\in G\setminus\{\emptyset\}}\tau(kw)\,\frac{1}{J}\sum_{j=1}^J\chi_j(w')\chi_j(w)=\tau(kw').
\end{equation}
If every entry of $\pi_\infty$ were equal to a common value, the first identity in \eqref{eq:s-orth} and the limit of \eqref{eq:s-invert} would give a zero limit for every $\tau(kw')$ with $w'\neq\emptyset$, which contradicts the assumption. Therefore, $\pi_\infty$ is not proportional to $\mathbf{1}$. By \eqref{eq:s-deltaclt}, which does not use Assumption \ref{ass:alias}, $\widehat\delta-\tau(k)\mathbf{1}-\pi=O_P(N^{-1/2})$. By \eqref{eq:s-uhat} and \eqref{eq:s-lambda}, which do not use Assumption \ref{ass:alias}, $N\widehat u_j\to4J\astar_j$ and $\widehat\lambda_j\to\lambda_j$ in probability. Every $\widehat\lambda_j$ lies in $[0,1]$, so $|\widehat\lambda^\top\mathsf{s}|\leq\max_j|\mathsf{s}_j|$ for every $\mathsf{s}\in\mathbb{R}^J$. Hence, for every $j$, since $\pi-\pi_\infty\to0$,
\begin{equation*}
    \widehat\delta_j-\tau(k)-\sum_{j'=1}^J\widehat\lambda_{j'}\{\widehat\delta_{j'}-\tau(k)\}=\pi_j-\widehat\lambda^\top\pi+O_P(N^{-1/2})=\pi_{\infty,j}-\lambda^\top\pi_\infty+o_P(1).
\end{equation*}
Dividing \eqref{eq:s-tshift} by $N$, with $N\widehat u_j=4J\astar_j+o_P(1)$ in the denominators by \eqref{eq:s-uhat},
\begin{equation}\label{eq:s-tlimit}
    \frac{\widehat T}{N}=\sum_{j=1}^J\frac{(\pi_{\infty,j}-\lambda^\top\pi_\infty)^2}{4J\astar_j}+o_P(1).
\end{equation}
The sum on the right, denoted by $\Theta_\infty$, is positive, because $\pi_\infty-(\lambda^\top\pi_\infty)\mathbf{1}$ is zero only when $\pi_\infty$ is proportional to $\mathbf{1}$. For $N\geq2\xi_\gamma/\Theta_\infty$, \eqref{eq:s-tlimit} gives
\begin{equation*}
    \bbP(\widehat T\leq\xi_\gamma)\leq\bbP\left(\frac{\widehat T}{N}\leq\frac{\Theta_\infty}{2}\right)\leq\bbP\left(\left|\frac{\widehat T}{N}-\Theta_\infty\right|\geq\frac{\Theta_\infty}{2}\right)\to0,
\end{equation*}
so $\bbP(\widehat T>\xi_\gamma)\to1$.

\section{Unequal allocations and staged assignment}\label{supp:unequal}

\subsection{Exact variances}\label{supp:unequal-exact}

\begin{proposition}\label{prop:s-unequal}
    Suppose that complete randomization assigns $n^{(j)}_z\geq2$ units to cell $z$ of $\cC_j$ under the fraction run on coset $j$, and $n_z\geq2$ units to cell $z$ of $\cZ$ under the full design, with totals $N$. Then the following hold for the unadjusted estimators.
    
    (i) The conclusion of Proposition \ref{prop:bias} of the main text holds.
    
    (ii) For every coset $j$,
    \begin{equation}\label{eq:s-uneq1}
        \begin{gathered}
        N\mathrm{var}_j(\widehat\tau_j)=\frac{1}{4^{K-p-1}}\sum_{z\in\cC_j}\frac{NS^2(z)}{n^{(j)}_z}-S^2\{\ttil_j(k)\},\quad N\bbE_j(\vhat_j)=\frac{1}{4^{K-p-1}}\sum_{z\in\cC_j}\frac{NS^2(z)}{n^{(j)}_z},
        \end{gathered}
    \end{equation}
    and
    \begin{equation}\label{eq:s-uneq2}
        \begin{gathered}
        N\mathrm{var}(\tauS)=\frac{1}{4^{K-1}}\sum_{z\in\cZ}\frac{NS^2(z)}{n_z}-S^2\{\tau(k)\},\quad N\bbE(\vhat_{\rm S})=\frac{1}{4^{K-1}}\sum_{z\in\cZ}\frac{NS^2(z)}{n_z}.
        \end{gathered}
    \end{equation}
    
    (iii) The average over cosets satisfies
    \begin{align*}
        \frac{1}{J}\sum_{j=1}^JN\mathrm{var}_j(\widehat\tau_j)-N\mathrm{var}(\tauS)&=\frac{1}{J}\sum_{j=1}^J\frac{1}{4^{K-p-1}}\sum_{z\in\cC_j}\frac{NS^2(z)}{n^{(j)}_z} \displaybreak[0]\\
        &\quad-\frac{1}{4^{K-1}}\sum_{z\in\cZ}\frac{NS^2(z)}{n_z}-\sum_{w\in G\setminus\{\emptyset\}}S^2\{\tau(kw)\}.
    \end{align*}
    
    (iv) The conclusion of part (iv) of Theorem \ref{thm:exact} of the main text holds, with the fraction on coset $j$ run with the cell sizes $n^{(j)}_z$.
\end{proposition}

\begin{proof}
    For part (i), the proof in \S\ref{supp:prop1} uses only the first identity in \eqref{eq:s-lidingvar}, which holds for any cell sizes $n_z\geq2$. For part (ii), \eqref{eq:s-lidingvar} with $c_z=2^{-(K-p-1)}g_k(z)$ on $\cC_j$ and with $c_z=2^{-(K-1)}g_k(z)$ on $\cZ$ gives the two variances, because the unit-level values of these contrasts are $\ttil_{j,i}(k)$ and $\tau_i(k)$. Since $\widehat\psi(z)$ is $s^2(z)$ divided by the size of cell $z$, \eqref{eq:s-s2unbiased} gives the two expectations. For part (iii), the identity \eqref{eq:s-hetavg} does not involve the allocation. Averaging the first identity in \eqref{eq:s-uneq1} over $j$, subtracting the first identity in \eqref{eq:s-uneq2}, and applying \eqref{eq:s-hetavg} proves the claim. For part (iv), the proof in \S\ref{supp:thm1} uses only Proposition \ref{prop:bias}, Lemma \ref{lem:alias}, the identities \eqref{eq:s-orth} and the law of total variance. None of these involves the allocation, and Proposition \ref{prop:bias} holds by part (i).
\end{proof}

The heterogeneity term in Proposition \ref{prop:s-unequal}(iii) is the same as in \eqref{eq:identity} of the main text. The allocation changes only the terms that the variance estimators estimate.

\begin{proposition}\label{prop:s-proportional}
    Suppose, in addition to the conditions of Proposition \ref{prop:s-unequal}, that $n^{(j)}_z=Jn_z$ for every $j$ and every $z\in\cC_j$. Then, for the unadjusted estimators, \eqref{eq:identity} of the main text holds and $J^{-1}\sum_{j=1}^JN\bbE_j(\vhat_j)=N\bbE(\vhat_{\rm S})$.
\end{proposition}

\begin{proof}
    Since $4^{K-1}=J^2\,4^{K-p-1}$,
    \begin{equation}\label{eq:s-prop1}
        \frac{1}{J}\sum_{j=1}^J\frac{1}{4^{K-p-1}}\sum_{z\in\cC_j}\frac{NS^2(z)}{Jn_z}=\frac{1}{J^2\,4^{K-p-1}}\sum_{z\in\cZ}\frac{NS^2(z)}{n_z}=\frac{1}{4^{K-1}}\sum_{z\in\cZ}\frac{NS^2(z)}{n_z}.
    \end{equation}
    Thus, the first two terms on the right-hand side of Proposition \ref{prop:s-unequal}(iii) cancel, which gives \eqref{eq:identity}. By the second identities in \eqref{eq:s-uneq1} and \eqref{eq:s-uneq2} with $n^{(j)}_z=Jn_z$, averaging $N\bbE_j(\vhat_j)$ over $j$ gives the left-hand side of \eqref{eq:s-prop1}, and $N\bbE(\vhat_{\rm S})$ is its right-hand side.
\end{proof}

The condition $n^{(j)}_z=Jn_z$ gives each cell of a fraction $J$ times as many units as the same cell of the full design, as in the balanced case. Since each design uses $N$ units, it implies $\sum_{z\in\cC_j}n_z=N/J$ for every $j$.

\subsection{Asymptotic results}\label{supp:unequal-asy}

Under $\cD$, let $e_z=n_z/N$ be fixed and positive for every $z\in\cZ$, and let
\begin{equation*}
    \mathsf{u}_j=\frac{1}{4^{K-p-1}}\sum_{z\in\cC_j}\frac{S^2_*(z)}{e_z},\qquad j=1,\ldots,J.
\end{equation*}
By \eqref{eq:s-psi}, which \S\ref{supp:limits} proves for any fixed positive allocation proportions, $\mathsf{u}_j$ is the limit in probability of $N\widehat u_j$. Let $\lambda^\dagger_j=\mathsf{u}_j^{-1}/\sum_{j'=1}^J\mathsf{u}_{j'}^{-1}$, which is well-defined because every $\mathsf{u}_j$ is positive by \eqref{eq:s-upos} below. When $e_z=2^{-K}$ for every $z$, $\mathsf{u}_j=4J\astar_j$ by \eqref{eq:s-uhat}, and $\lambda^\dagger_j=\lambda_j$. Under $\cD_j$, let $e^{(j)}_z=n^{(j)}_z/N$ be fixed and positive for every $z\in\cC_j$, and let
\begin{equation*}
    \mathsf{v}_j=\frac{1}{4^{K-p-1}}\sum_{z\in\cC_j}\frac{S^2_*(z)}{e^{(j)}_z}.
\end{equation*}
When $e^{(j)}_z=2^{-(K-p)}$ for every $z$, $\mathsf{v}_j=4\astar_j$ by \eqref{eq:s-varj}. Here, $N$ runs through values for which every $e_zN$ and every $e^{(j)}_zN$ is an integer.

\begin{proposition}\label{prop:s-unequal-asy}
    Suppose that under $\cD$ complete randomization assigns $n_z\geq2$ units to cell $z$, with $n_z/N=e_z$ fixed and positive, and that the conditions of Theorem \ref{thm:asy} of the main text hold. Then the following hold.
    
    (i) $N^{1/2}\{\tauS-\tau(k)\}\to\mathcal{N}[0,\,J^{-2}\sum_{j=1}^J\mathsf{u}_j-S^2\{\tau^*(k)\}]$ in distribution, $N\vhat_{\rm S}\to J^{-2}\sum_{j=1}^J\mathsf{u}_j$ and $N\vhat_{\rm W}\to(\sum_{j=1}^J\mathsf{u}_j^{-1})^{-1}$ in probability, and $(\sum_{j=1}^J\mathsf{u}_j^{-1})^{-1}\leq J^{-2}\sum_{j=1}^J\mathsf{u}_j$, with equality if and only if $\mathsf{u}_1=\cdots=\mathsf{u}_J$. Also, $\vhat_{\rm W}\leq\vhat_{\rm S}$ for every assignment.
    
    (ii) If also Assumption \ref{ass:alias} holds, then $N^{1/2}\{\tauW-\tau(k)\}\to\mathcal{N}[0,\,(\sum_j\mathsf{u}_j^{-1})^{-1}-S^2\{\sum_j\lambda^\dagger_j\ttil^*_j(k)\}]$ in distribution.
    
    (iii) The conclusions of Proposition \ref{prop:test} of the main text hold.
    
    (iv) If under $\cD_j$ complete randomization assigns $n^{(j)}_z\geq2$ units to cell $z\in\cC_j$, with $n^{(j)}_z/N=e^{(j)}_z$ fixed and positive, then under $\cD_j$, $N^{1/2}\{\widehat\tau_j-\ttil_j(k)\}\to\mathcal{N}[0,\,\mathsf{v}_j-S^2\{\ttil^*_j(k)\}]$ in distribution and $N\vhat_j\to \mathsf{v}_j$ in probability.
\end{proposition}

\begin{proof}
    By \S\ref{supp:limits}, \eqref{eq:s-clt} and \eqref{eq:s-psi} hold for these fixed positive proportions. By \eqref{eq:s-psi},
    \begin{equation}\label{eq:s-uhat-gen}
        N\widehat u_j=\frac{1}{4^{K-p-1}}\sum_{z\in\cC_j}N\widehat\psi(z)=\mathsf{u}_j+o_P(1),\qquad j=1,\ldots,J.
    \end{equation}
    Since $e_z\leq1$ and $\sum_{z\in\cC_j}S^2_*(z)=2^{K-p}\astar_j$,
    \begin{equation}\label{eq:s-upos}
        \mathsf{u}_j\geq\frac{2^{K-p}}{4^{K-p-1}}\,\astar_j,
    \end{equation}
    and the limit of $\astar_j$ is positive by Assumption \ref{ass:limits} or \ref{ass:cov}. For part (i), take $c_z=2^{-(K-1)}g_k(z)$ in \eqref{eq:s-clt}. Then, $\sum_zc_z\Ystar_i(z)=\tau^*_i(k)$, and $4^{K-1}=J^2\,4^{K-p-1}$ gives $\sum_{z\in\cZ}c_z^2S^2_*(z)/e_z=4^{-(K-1)}\sum_{z\in\cZ}S^2_*(z)/e_z=J^{-2}\sum_j\mathsf{u}_j$, which gives the limit for $\tauS$. The same identity gives $\vhat_{\rm S}=4^{-(K-1)}\sum_{z\in\cZ}\widehat\psi(z)=J^{-2}\sum_j\widehat u_j$, so \eqref{eq:s-uhat-gen} gives $N\vhat_{\rm S}=J^{-2}\sum_j\mathsf{u}_j+o_P(1)$. By \eqref{eq:s-uhat-gen} and \eqref{eq:s-upos},
    \begin{equation*}
        \bbP\left(\min_{1\leq j\leq J}\widehat u_j\leq0\right)\leq\sum_{j=1}^J\bbP\left(|N\widehat u_j-\mathsf{u}_j|\geq \mathsf{u}_j\right)\to0.
    \end{equation*}
    On the event that every $\widehat u_j>0$, and by \eqref{eq:s-event} for that event and the continuous mapping theorem,
    \begin{equation*}
        N\vhat_{\rm W}=\left\{\sum_{j=1}^J(N\widehat u_j)^{-1}\right\}^{-1}=\left(\sum_{j=1}^J\mathsf{u}_j^{-1}\right)^{-1}+o_P(1).
    \end{equation*}
    The Cauchy--Schwarz inequality gives $J^2=\{\sum_j\mathsf{u}_j^{1/2}\mathsf{u}_j^{-1/2}\}^2\leq\sum_j\mathsf{u}_j\sum_j\mathsf{u}_j^{-1}$, which rearranges to the stated inequality, with equality if and only if $\mathsf{u}_1=\cdots=\mathsf{u}_J$. The identity $\vhat_{\rm S}=J^{-2}\sum_j\widehat u_j$ holds for every assignment. When every $\widehat u_j>0$, $\vhat_{\rm W}=(\sum_j\widehat u_j^{-1})^{-1}$, and the same inequality with $\widehat u_j$ in place of $\mathsf{u}_j$ gives $\vhat_{\rm W}\leq\vhat_{\rm S}$. Otherwise $\widehat\lambda_j=1/J$ for every $j$, and $\vhat_{\rm W}=\vhat_{\rm S}$.
    
    For part (ii), \eqref{eq:s-deltaclt} holds because \eqref{eq:s-clt} does. On the event that every $\widehat u_j>0$, whose probability tends to one by part (i), $\widehat\lambda_j=(N\widehat u_j)^{-1}/\sum_{j'}(N\widehat u_{j'})^{-1}$. By \eqref{eq:s-event} for that event, \eqref{eq:s-uhat-gen} and the continuous mapping theorem, $\widehat\lambda_j-\lambda^\dagger_j=o_P(1)$. Under Assumption \ref{ass:alias}, $\ttil_j(k)=\tau(k)$ for every $j$, and \eqref{eq:s-decomp} holds with $\lambda_j$ replaced by $\lambda^\dagger_j$. Its second sum is $o_P(1)\,O_P(1)=o_P(1)$ by $\widehat\lambda_j-\lambda^\dagger_j=o_P(1)$ and \eqref{eq:s-deltaclt}. Its first sum is the left-hand side of \eqref{eq:s-clt} with $c_z=\lambda^\dagger_{j(z)}2^{-(K-p-1)}g_k(z)$, for which $\sum_zc_z\Ystar_i(z)=\sum_j\lambda^\dagger_j\ttil^*_{j,i}(k)$ and
    \begin{equation*}
        \sum_{z\in\cZ}\frac{c_z^2S^2_*(z)}{e_z}=\sum_{j=1}^J(\lambda^\dagger_j)^2\,\frac{1}{4^{K-p-1}}\sum_{z\in\cC_j}\frac{S^2_*(z)}{e_z}=\sum_{j=1}^J(\lambda^\dagger_j)^2\mathsf{u}_j=\left(\sum_{j=1}^J\mathsf{u}_j^{-1}\right)^{-1}.
    \end{equation*}
    Slutsky's theorem gives the stated limit.
    
    For part (iii), take $c_z=\mathsf{s}_{j(z)}2^{-(K-p-1)}g_k(z)$ for a fixed $\mathsf{s}\in\mathbb{R}^J$, as in \S\ref{supp:test}. Then,
    \begin{equation*}
        \sum_{z\in\cZ}\frac{c_z^2S^2_*(z)}{e_z}=\sum_{j=1}^J\mathsf{s}_j^2\mathsf{u}_j,
    \end{equation*}
    and $S^2\{\sum_zc_z\Ystar(z)\}=\mathsf{s}^\top\Sigma_*\mathsf{s}$ does not depend on the allocation. Thus, \eqref{eq:s-vecclt} holds with $D=\mathrm{diag}(\mathsf{u}_1,\ldots,\mathsf{u}_J)$, and \eqref{eq:s-uhat-gen} gives $\widehat D-D=o_P(1)$. By \eqref{eq:s-upos}, the diagonal entries of $D$ are positive. The proof in \S\ref{supp:test} uses only these three facts about $D$, together with the convergence of $\widehat\lambda_j$ in probability to the $j$th entry of $D^{-1}\mathbf{1}/(\mathbf{1}^\top D^{-1}\mathbf{1})$. Here, that entry is $\lambda^\dagger_j$, and part (ii) shows the convergence. Thus, the same proof gives the conclusions of Proposition \ref{prop:test} under these allocation proportions, with $4J\astar_j$ replaced by $\mathsf{u}_j$ and $\lambda$ by $\lambda^\dagger=(\lambda^\dagger_1,\ldots,\lambda^\dagger_J)^\top$ in \eqref{eq:s-tlimit}.
    
    For part (iv), \eqref{eq:s-clt} and \eqref{eq:s-psi} hold under $\cD_j$ with $\cQ=\cC_j$ and the fixed proportions $e^{(j)}_z$. With $c_z=2^{-(K-p-1)}g_k(z)$, $\sum_zc_z\overline Y(z)=\ttil_j(k)$, $\sum_zc_z\Ystar_i(z)=\ttil^*_{j,i}(k)$ and
    \begin{equation*}
        \sum_{z\in\cC_j}\frac{c_z^2S^2_*(z)}{e^{(j)}_z}=\mathsf{v}_j,\qquad N\vhat_j=\frac{1}{4^{K-p-1}}\sum_{z\in\cC_j}N\widehat\psi(z)=\mathsf{v}_j+o_P(1),
    \end{equation*}
    where the second identity is by \eqref{eq:s-psi}. Therefore, \eqref{eq:s-clt} gives the stated limit.
\end{proof}

When $n^{(j)}_z=Jn_z$ for every $j$ and $z\in\cC_j$, $e^{(j)}_z=Je_z$ and $\mathsf{v}_j=\mathsf{u}_j/J$. The average of the limits $\mathsf{v}_j$ of $N\vhat_j$ over $j$ is then $J^{-2}\sum_j\mathsf{u}_j$, the limit of $N\vhat_{\rm S}$. Thus, part (i) of Proposition \ref{prop:s-unequal-asy} compares $\tauW$ with the average fraction as Theorem \ref{thm:asy}(iii) does. Under unequal allocations, the weights $\lambda^\dagger_j$ and the condition for equality in part (i) depend on the $\mathsf{u}_j$, which weight each cell variance by $1/e_z$, rather than on the average cell variances.

\begin{corollary}\label{cor:s-unequal}
    Suppose that the conditions of Proposition \ref{prop:s-unequal-asy} and Assumption \ref{ass:additive} of the main text hold. Then, $S^2\{\ttil^*_j(k)\}=S^2\{\sum_{j'}\lambda^\dagger_{j'}\ttil^*_{j'}(k)\}=S^2\{\tau^*(k)\}$ for every $j$, and
    \begin{equation*}
        \frac{1}{J^2}\sum_{j=1}^J\mathsf{u}_j-S^2\{\tau^*(k)\}\geq\left(\sum_{j=1}^J\mathsf{u}_j^{-1}\right)^{-1}-S^2\left\{\sum_{j=1}^J\lambda^\dagger_j\ttil^*_j(k)\right\},
    \end{equation*}
    with equality in the inequality if and only if $\mathsf{u}_1=\cdots=\mathsf{u}_J$. If also $n^{(j)}_z=Jn_z$ for every $j$ and $z\in\cC_j$, then $J^{-1}\sum_{j=1}^J[\mathsf{v}_j-S^2\{\ttil^*_j(k)\}]=J^{-2}\sum_{j=1}^J\mathsf{u}_j-S^2\{\tau^*(k)\}$.
\end{corollary}

\begin{proof}
    By Assumption \ref{ass:additive}, $\tau^*_i(kw)=\tau^*(kw)$ for every unit $i$ and every $w\in G\setminus\{\emptyset\}$. Lemma \ref{lem:alias} applied to $\Ystar$ gives $\ttil^*_{j,i}(k)-\tau^*_i(k)=\sum_{w\in G\setminus\{\emptyset\}}\chi_j(w)\tau^*(kw)$, which does not depend on $i$. Since $\sum_j\lambda^\dagger_j=1$, also $\sum_j\lambda^\dagger_j\ttil^*_{j,i}(k)-\tau^*_i(k)=\sum_j\lambda^\dagger_j\sum_{w\in G\setminus\{\emptyset\}}\chi_j(w)\tau^*(kw)$, which does not depend on $i$ either. Adding the same constant to the value of every unit does not change $S^2(\cdot)$, so $S^2\{\ttil^*_j(k)\}=S^2\{\sum_{j'}\lambda^\dagger_{j'}\ttil^*_{j'}(k)\}=S^2\{\tau^*(k)\}$ for every $j$. The inequality and its equality case are part (i) of Proposition \ref{prop:s-unequal-asy}, from which the common term $S^2\{\tau^*(k)\}$ is subtracted. Under $n^{(j)}_z=Jn_z$, $\mathsf{v}_j=\mathsf{u}_j/J$, and averaging $\mathsf{v}_j-S^2\{\tau^*(k)\}$ over $j$ gives the last identity.
\end{proof}

The left-hand side of the inequality in Corollary \ref{cor:s-unequal} is the asymptotic variance of $\tauS$, by part (i) of Proposition \ref{prop:s-unequal-asy}. Under Assumption \ref{ass:alias}, the right-hand side is that of $\tauW$, by part (ii). The left-hand side of its last identity is the average over cosets of the asymptotic variances of the fractions, by part (iv).

\subsection{Staged assignment}\label{supp:staged}

Consider a set $\mathcal{J}\subseteq\{1,\ldots,J\}$ of cosets and the cells $\cQ=\bigcup_{j\in\mathcal{J}}\cC_j$, with sizes $n_z\geq2$ that sum to $N$, and let $N_j=\sum_{z\in\cC_j}n_z$. The two-stage assignment runs the cosets in $\mathcal{J}$ in separate stages. First, the $N$ units are split completely at random into labeled groups, with $N_j$ units in group $j\in\mathcal{J}$. Second, independently across groups, the units of group $j$ are assigned by complete randomization to the cells of $\cC_j$, with $n_z$ units in cell $z$.

\begin{lemma}\label{lem:s-staged}
    The two-stage assignment and complete randomization of the $N$ units to the cells of $\cQ$, with $n_z$ units in cell $z$, have the same distribution of the assignment.
\end{lemma}

\begin{proof}
    Under complete randomization, every assignment with $n_z$ units in each cell $z\in\cQ$ has probability $\prod_{z\in\cQ}n_z!/N!$. Fix such an assignment, and let $\mathcal{U}_j$ be the set of units it places in the cells of $\cC_j$, which has $N_j$ elements. The two-stage assignment produces it only if the first stage forms the groups $\mathcal{U}_j$, $j\in\mathcal{J}$, which has probability $\prod_{j\in\mathcal{J}}N_j!/N!$. Given these groups, the second stage produces it with probability $\prod_{j\in\mathcal{J}}\{\prod_{z\in\cC_j}n_z!/N_j!\}$. The product of the two probabilities is $\prod_{z\in\cQ}n_z!/N!$.
\end{proof}

With $\mathcal{J}=\{1,\ldots,J\}$, complete randomization to the cells of $\cQ$ is $\cD$ with cell sizes $n_z$. Thus, running all cosets in stages leaves every result for $\cD$ unchanged. With $\mathcal{J}=\{j,j'\}$ and $p\geq2$, $\cQ$ is a coset of a regular fraction by Lemma \ref{lem:s-union}. Therefore, a fold-over that runs coset $j$ and then coset $j'$ in a second stage is that fraction. The results for fractions apply to it with $G$ replaced by the subgroup of Lemma \ref{lem:s-union}, and Propositions \ref{prop:s-unequal} and \ref{prop:s-unequal-asy}(iv) apply when the two stages have different sizes. Both statements require that every later stage is planned before the first is run, and that the potential outcomes do not depend on the stage. A fold-over chosen after the first stage has been analyzed, or run on new units, is not covered. When the stages are instead fixed groups of units, such as successive batches, the cosets become strata, and Proposition \ref{prop:bias} and the variances of \S\ref{supp:exact} no longer apply.

\section{Simulation studies}\label{supp:sim}

\subsection{Three scenarios}\label{supp:sim-table}

This study checks Theorems \ref{thm:exact} and \ref{thm:asy} of the main text in three scenarios that separate the two sources of difference between the designs. There are $N=400$ units and $K=3$ factors. The generator is $123$, and the target is the main effect of factor 1, which is aliased with the interaction $23$. The potential outcomes are generated as
\begin{equation*}
    Y_i(z)=1+\alpha_i+(1+\nu_i)\frac{z_1}{2}+\frac{z_2+z_3}{4}+\eta\,\omega_iz_2z_3+\kappa\,\alpha_iz_1z_2z_3,
\end{equation*}
with $\alpha_i=0.8x_i+0.6\varepsilon_i$, where $x_i$, $\varepsilon_i$, $\nu_i$ and $\omega_i$ are independent standard normal draws, each centered and scaled to unit finite-population variance, and fixed across replications. Then, $\tau(1)=1$, $\tau(23)=0$, and $\tau_i(23)=2\eta\,\omega_i$, so that $S^2\{\tau(23)\}=4\eta^2$. Scenario 1 has $\eta=\kappa=0$. Scenario 2 has $\eta=1$ and $\kappa=0$, so the aliased effect varies across units. Scenario 3 has $\eta=0$ and $\kappa=0.4$, and thus the two cosets have different cell variances. Each replication draws one assignment from each of $\cD$, $\cD_1$ and $\cD_2$, and computes the unadjusted versions in Definition \ref{def:est} of the main text and their $95\%$ Wald intervals, over $1{,}000$ replications. The Monte Carlo standard error of $N\mathrm{var}$ is about $4.5\%$ of the Monte Carlo value, and that of the coverage is at most $0.7$ percentage points. Table \ref{tab:sim} of the main text reports $N$ times the variance from Theorems \ref{thm:exact} and \ref{thm:asy}, $N$ times the Monte Carlo variance, $N$ times the mean estimated variance, and the coverage.

The results match Theorems \ref{thm:exact} and \ref{thm:asy} of the main text. In Scenario 1, all four procedures have the same theoretical variance and the same mean estimated variance. In Scenario 2, the fractions have a smaller variance than the full design, as predicted by \eqref{eq:identity} of the main text, but all four mean estimated variances are about the same. Therefore, the intervals of the fractions over-cover, while those of $\tauS$ and $\tauW$ cover close to the nominal level. In Scenario 3, the mean estimated variances of the two fractions average to that of $\tauS$, as predicted by Theorem \ref{thm:exact}(iii), and $\tauW$ has a smaller mean estimated variance than $\tauS$. The Monte Carlo variances are within three Monte Carlo standard errors of their theoretical values.

\subsection{Two sweeps}\label{supp:sim-figure}

\begin{figure}[!tbp]
    \centering
    \includegraphics[width=\textwidth]{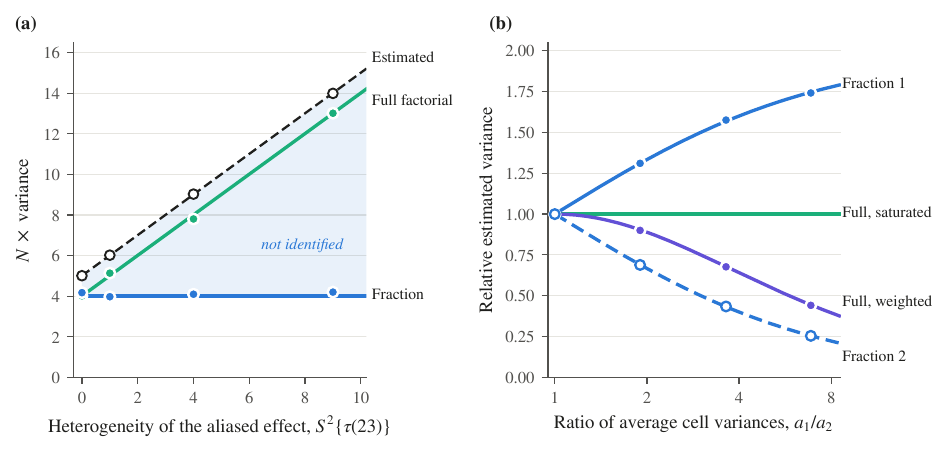}
    \caption{(a) $N$ times the variance of the fraction estimator averaged over the two cosets (blue), $N$ times the variance of $\tauS$ (green), and the expectation of $N$ times the estimated variance of $\tauS$, which equals the average over the two fractions (black, dashed), against $S^2\{\tau(23)\}$. The shaded band lies between the variance of the fraction and its estimated variance. Filled circles are Monte Carlo variances, and open circles are Monte Carlo means of $N\vhat$ for the fractions. (b) Estimated variance relative to that of $\tauS$ for $\widehat\tau_1$ (blue, solid), $\widehat\tau_2$ (blue, dashed, open circles), $\tauS$ (green) and $\tauW$ (violet), against $a_1/a_2$ on a logarithmic scale. Circles are Monte Carlo values. Lines are the exact values from Theorem \ref{thm:exact} in panel (a) and the limits implied by Theorems \ref{thm:exact} and \ref{thm:asy} in panel (b), and the Monte Carlo values use $N=400$ over $1{,}000$ replications.}
    \label{fig:s-sweep}
\end{figure}

This study traces the two terms in Theorem \ref{thm:exact} of the main text as the scenario parameters vary. The design and estimators are the same as in \S\ref{supp:sim-table}. Panel (a) of Figure \ref{fig:s-sweep} varies $\eta$ over $0$, $0.5$, $1$ and $1.5$ with $\kappa=0$, and panel (b) varies $\kappa$ over $0$, $0.2$, $0.4$ and $0.6$ with $\eta=0$, with $1{,}000$ replications at each value. Figure \ref{fig:s-sweep} reports the true and estimated variances against the heterogeneity of the aliased effect in panel (a), and the estimated variances relative to that of $\tauS$ against the ratio $\rho=a_1/a_2$ in panel (b).

The results match Theorems \ref{thm:exact} and \ref{thm:asy} of the main text. In panel (a), the variance of the fraction does not change as the aliased effect becomes more heterogeneous. The variance of $\tauS$ and the estimated variances of both designs grow with $S^2\{\tau(23)\}$, so the gain that the variance estimator misses grows as well. In panel (b), the estimated variances of the two fractions diverge as the cosets become unequal, while their average remains equal to that of $\tauS$. The weighted estimator lies between the better fraction and $\tauS$, and its advantage over $\tauS$ grows with $\rho$. The Monte Carlo values agree with the theoretical curves within Monte Carlo error.

\FloatBarrier

\subsection{Adjusted estimators}\label{supp:sim-lin}

This study checks Theorem \ref{thm:asy} for the adjusted estimators of Definition \ref{def:est}. The design, the three scenarios, and the number of replications are the same as in \S\ref{supp:sim-table}, and the covariate $x_i$ is correlated with $\alpha_i$ at $0.8$. Table \ref{tab:sim-lin} reports $N$ times the asymptotic variance with the residual potential outcomes, $N$ times the Monte Carlo variance, $N$ times the mean estimated variance, and the coverage.

\begin{table}[htbp]
    \caption{Simulation results with the adjusted versions in Definition \ref{def:est} of the main text, Scenarios 1--3. The theory column is the asymptotic variance of Theorem \ref{thm:asy} evaluated at the finite population with the residual potential outcomes. Other columns are the same as in Table \ref{tab:sim} of the main text.}
    \centering
    \begin{tabular}{ccrrrr}
        \toprule
        Scenario & Estimator & \hd{$N\mathrm{var}$}{(theory)} & \hd{$N\mathrm{var}$}{(Monte Carlo)} & \hd{$N\times$ mean}{$\widehat v$} & \hd{Coverage}{(\%)} \\
        \midrule
        1 & $\widehat\tau_1$ & 1.44 & 1.46 & 2.41 & 98.5 \\
         & $\widehat\tau_2$ & 1.44 & 1.42 & 2.42 & 98.9 \\
         & $\widehat\tau_{\rm S}$ & 1.44 & 1.49 & 2.40 & 99.4 \\
         & $\widehat\tau_{\rm W}$ & 1.44 & 1.51 & 2.38 & 99.1 \\
        2 & $\widehat\tau_1$ & 1.44 & 1.44 & 6.31 & 100.0 \\
         & $\widehat\tau_2$ & 1.44 & 1.55 & 6.37 & 100.0 \\
         & $\widehat\tau_{\rm S}$ & 5.43 & 5.61 & 6.28 & 96.9 \\
         & $\widehat\tau_{\rm W}$ & 5.43 & 5.60 & 6.25 & 96.8 \\
        3 & $\widehat\tau_1$ & 2.82 & 2.94 & 3.80 & 97.0 \\
         & $\widehat\tau_2$ & 0.52 & 0.50 & 1.50 & 100.0 \\
         & $\widehat\tau_{\rm S}$ & 1.67 & 1.81 & 2.61 & 98.1 \\
         & $\widehat\tau_{\rm W}$ & 1.17 & 1.28 & 2.12 & 98.5 \\
    \bottomrule
    \end{tabular}
    \label{tab:sim-lin}
\end{table}

The results match Theorem \ref{thm:asy} of the main text with the residual potential outcomes in place of the potential outcomes. Adjustment removes the part of $\alpha_i$ explained by $x_i$, so every variance and every mean estimated variance is smaller than in Table \ref{tab:sim} of the main text. Otherwise, the pattern remains the same as in Table \ref{tab:sim}. In Scenario 2, the fractions have a smaller asymptotic variance than the full design, but all four mean estimated variances are about the same. In Scenario 3, the mean estimated variances of the two fractions average close to that of $\tauS$, and $\tauW$ has a smaller mean estimated variance than $\tauS$.

\subsection{A nonzero aliased effect}\label{supp:sim-fail}

This study checks Proposition \ref{prop:bias} when Assumption \ref{ass:alias} fails. The design and the number of replications are the same as in \S\ref{supp:sim-table}, with $\eta=\kappa=0$ and the term $z_2z_3/4$ added to the potential outcomes, so that $\tau(23)=0.5$. We call this Scenario 4. Both versions of the estimators are computed. Table \ref{tab:sim-fail} reports the bias, $N$ times the theoretical and Monte Carlo variances, $N$ times the mean estimated variance, and the coverage.

\begin{table}[htbp]
    \caption{Simulation results for Scenario 4, in which $\tau(23)=0.5$ so that Assumption \ref{ass:alias} of the main text fails. The first block is unadjusted and the second is adjusted. Bias is the Monte Carlo mean minus $\tau(1)=1$. The theory column is Theorem \ref{thm:exact}(i) of the main text for the unadjusted versions and the asymptotic variance of Theorem \ref{thm:asy}(i) with the residual potential outcomes for the adjusted ones. No theoretical variance is given for $\widehat\tau_{\rm W}$, to which Theorem \ref{thm:asy}(ii) does not apply. Other columns are the same as in Table \ref{tab:sim} of the main text.}
    \centering
    \begin{tabular}{ccrrrrr}
        \toprule
        Version & Estimator & Bias & \hd{$N\mathrm{var}$}{(theory)} & \hd{$N\mathrm{var}$}{(Monte Carlo)} & \hd{$N\times$ mean}{$\widehat v$} & \hd{Coverage}{(\%)} \\
        \midrule
        Unadjusted & $\widehat\tau_1$ & 0.502 & 4.01 & 4.36 & 5.00 & 0.5 \\
         & $\widehat\tau_2$ & $-$0.498 & 4.01 & 3.86 & 5.02 & 0.1 \\
         & $\widehat\tau_{\rm S}$ & $-$0.004 & 4.01 & 4.19 & 5.01 & 97.1 \\
         & $\widehat\tau_{\rm W}$ & $-$0.005 & -- & 4.82 & 4.99 & 95.3 \\
        Adjusted & $\widehat\tau_1$ & 0.497 & 1.44 & 1.37 & 2.42 & 0.0 \\
         & $\widehat\tau_2$ & $-$0.498 & 1.44 & 1.47 & 2.40 & 0.0 \\
         & $\widehat\tau_{\rm S}$ & 0.002 & 1.44 & 1.54 & 2.39 & 98.5 \\
         & $\widehat\tau_{\rm W}$ & 0.002 & -- & 2.14 & 2.38 & 95.7 \\
    \bottomrule
    \end{tabular}
    \label{tab:sim-fail}
\end{table}

The results match Proposition \ref{prop:bias} of the main text for the unadjusted versions and Theorem \ref{thm:asy}(i) for the adjusted ones. Each fraction is biased by the aliased effect, with the sign of its coset, and its interval almost never covers $\tau(1)$. The saturated estimator is unbiased, and its interval maintains its coverage. The weighted estimator is nearly unbiased here, because the two cosets have equal average cell variances and its weights are close to $1/2$. Theorem \ref{thm:asy}(ii) does not apply to it, and its variance is larger than that of $\tauS$. With two cosets, $\tauW-\tauS=(\widehat\lambda_1-1/2)(\widehat\delta_1-\widehat\delta_2)$, and $\widehat\delta_1-\widehat\delta_2$ tends to $2\tau(23)$ rather than to zero. Thus, the random weight increases variance.

\subsection{A test of the aliasing assumption}\label{supp:sim-test}

This study checks Proposition \ref{prop:test} of the main text. The design, the four scenarios, and the number of replications are the same as in \S\ref{supp:sim-table} and \S\ref{supp:sim-fail}, and the test rejects when $\widehat T$ exceeds the $0.95$-quantile of the chi-squared distribution with one degree of freedom. Both versions of the estimators are used. With $J=2$, $\widehat T=(\widehat\delta_1-\widehat\delta_2)^2/(\widehat u_1+\widehat u_2)$. Under Assumption \ref{ass:alias}, \eqref{eq:s-vecclt} with $\mathsf{s}=(1,-1)^\top$ gives $N^{1/2}(\widehat\delta_1-\widehat\delta_2)\to\mathcal{N}[0,\,4J(\astar_1+\astar_2)-S^2\{\ttil^*_1(k)-\ttil^*_2(k)\}]$ in distribution, and \eqref{eq:s-uhat} gives $N(\widehat u_1+\widehat u_2)\to4J(\astar_1+\astar_2)$ in probability. Thus, $\widehat T$ converges in distribution to $\mathsf{q}$ times a chi-squared variable with one degree of freedom, where $\mathsf{q}=1-S^2\{\ttil^*_1(k)-\ttil^*_2(k)\}/\{4J(\astar_1+\astar_2)\}$. Table \ref{tab:test} reports the rejection rates and, for Scenarios 1--3, the limiting rejection rates with $\mathsf{q}$ evaluated at the finite population.

\begin{table}[htbp]
    \caption{Rejection rates (\%) of the test of Proposition \ref{prop:test} of the main text at level $0.05$, with $N=400$ over $1{,}000$ replications. Assumption \ref{ass:alias} holds in Scenarios 1--3 and fails in Scenario 4. The limit is the limiting rejection rate from the proof of Proposition \ref{prop:test} with the finite-population values, and is not given for Scenario 4, where it is $100\%$. The Monte Carlo standard error of a rate near $5\%$ is $0.7$ percentage points.}
    \centering
    \begin{tabular}{crrrr}
        \toprule
        & \multicolumn{2}{c}{Unadjusted} & \multicolumn{2}{c}{Adjusted} \\
        \cmidrule(lr){2-3}\cmidrule(lr){4-5}
        Scenario & Rate & Limit & Rate & Limit \\
        \midrule
        1 & 4.6 & 5.0 & 6.2 & 5.0 \\
        2 & 0.9 & 0.9 & 0.2 & 0.1 \\
        3 & 5.0 & 5.0 & 6.0 & 5.0 \\
        4 & 99.6 & -- & 100.0 & -- \\
    \bottomrule
    \end{tabular}
    \label{tab:test}
\end{table}

The results match Proposition \ref{prop:test} of the main text. In Scenarios 1 and 3, every aliased effect is zero for every unit, and the unadjusted test rejects at about the nominal rate, as predicted by part (i). The adjusted test rejects slightly more often, within two Monte Carlo standard errors of the nominal rate. In Scenario 2, the aliased effect varies across units, and the test is conservative, with rejection rates close to their limits. In Scenario 4, the test almost always rejects, as predicted by part (ii).

\subsection{Four cosets}\label{supp:sim-j4}

This study checks Theorem \ref{thm:asy} and Proposition \ref{prop:test} of the main text with four cosets. There are $K=5$ factors. The generators are $124$ and $135$, so that $J=4$, and the target is the main effect of factor 1, which is aliased with $24$, $35$, and $12345$. The potential outcomes are generated as
\begin{align*}
    Y_i(z)=&~1+\alpha_i+(1+\nu_i)\frac{z_1}{2}+\frac{z_2+z_3+z_4+z_5}{4}+\left\{\eta\,\omega_i+\frac{\tau(24)}{2}\right\}z_2z_4\displaybreak[0]\\
    &~+\alpha_i(\kappa_1z_1z_2z_4+\kappa_2z_1z_3z_5),
\end{align*}
with $\alpha_i$, $x_i$, $\nu_i$ and $\omega_i$ the same as in \S\ref{supp:sim-table}, fixed across replications. Scenario A has $\eta=\tau(24)=0$ and $(\kappa_1,\kappa_2)=(0.4,0.2)$, so Assumptions \ref{ass:alias} and \ref{ass:additive} hold and the four cosets have different average cell variances. Scenario B has $\eta=1$ and $\kappa_1=\kappa_2=\tau(24)=0$, with the aliased effect $\tau_i(24)=2\omega_i$ varying across units. Scenario C has $\tau(24)=0.5$ and $\eta=\kappa_1=\kappa_2=0$, and thus Assumption \ref{ass:alias} fails. There are $N=640$ or $N=2{,}560$ units, so that $\cD$ has $20$ or $80$ units in each cell. Each of the $2{,}000$ replications draws one assignment from each of $\cD,\cD_1,\ldots,\cD_4$, and both versions of the estimators are used. The test rejects when $\widehat T$ exceeds the $0.95$-quantile of the chi-squared distribution with three degrees of freedom. Table \ref{tab:j4} reports, for $\tauS$ and $\tauW$, $N$ times the theoretical and Monte Carlo variances, $N$ times the mean estimated variance and the coverage, together with the coverage of the fractions averaged over the four cosets and the rejection rate of the test. We use $2{,}000$ replications so that the Monte Carlo standard error of a rejection rate near $5\%$ is about $0.5$ percentage points.

\begin{table}[htbp]
    \caption{Simulation results with four cosets, $K=5$, generators $124$ and $135$, and target $\tau(1)$, over $2{,}000$ replications. For $\widehat\tau_{\rm S}$ and $\widehat\tau_{\rm W}$, the columns are $N\mathrm{var}$ from Theorem \ref{thm:exact}(i) or Theorem \ref{thm:asy}, as in Table \ref{tab:sim} of the main text, $N\mathrm{var}$ from Monte Carlo, $N$ times the mean estimated variance, and the coverage (\%). The theory column for $\widehat\tau_{\rm W}$ is not given in Scenario C, where Assumption \ref{ass:alias} fails. Fractions is the coverage (\%) of $\tau(1)$ averaged over the four fractions, and Reject is the rejection rate (\%) of the test of Proposition \ref{prop:test} at level $0.05$. The Monte Carlo standard error of $N\mathrm{var}$ is about $3.2\%$ of the Monte Carlo value, that of a coverage near $95\%$ is $0.5$ percentage points, and that of a rejection rate near $5\%$ is $0.5$ percentage points.}
    \centering
    \resizebox{\textwidth}{!}{
    \begin{tabular}{cccrrrrrrrrrr}
        \toprule
        & & & \multicolumn{4}{c}{$\widehat\tau_{\rm S}$} & \multicolumn{4}{c}{$\widehat\tau_{\rm W}$} & & \\
        \cmidrule(lr){4-7}\cmidrule(lr){8-11}
        Version & Scenario & $N$ & Theory & MC & $\widehat v$ & Cov. & Theory & MC & $\widehat v$ & Cov. & Fractions & Reject \\
        \midrule
        Unadjusted & A & 640 & 5.02 & 5.00 & 6.03 & 96.4 & 2.64 & 2.67 & 3.61 & 97.6 & 97.5 & 6.2 \\
         & A & 2560 & 4.66 & 4.43 & 5.66 & 97.1 & 2.49 & 2.39 & 3.48 & 98.6 & 97.6 & 4.4 \\
         & B & 640 & 8.18 & 8.16 & 9.20 & 95.3 & 8.18 & 8.24 & 9.09 & 95.4 & 99.6 & 3.4 \\
         & B & 2560 & 7.89 & 7.82 & 8.89 & 96.3 & 7.89 & 7.83 & 8.87 & 96.4 & 99.6 & 2.7 \\
         & C & 640 & 4.18 & 4.31 & 5.18 & 96.5 & -- & 5.11 & 5.12 & 94.7 & 0.0 & 100.0 \\
         & C & 2560 & 3.89 & 3.91 & 4.89 & 96.4 & -- & 4.42 & 4.88 & 96.3 & 0.0 & 100.0 \\
        Adjusted & A & 640 & 1.72 & 1.84 & 2.58 & 98.2 & 1.13 & 1.23 & 2.00 & 99.1 & 98.6 & 8.7 \\
         & A & 2560 & 1.73 & 1.78 & 2.69 & 98.6 & 1.14 & 1.13 & 2.10 & 99.3 & 98.7 & 6.2 \\
         & B & 640 & 5.43 & 5.81 & 6.06 & 95.5 & 5.42 & 5.85 & 5.98 & 95.3 & 100.0 & 4.9 \\
         & B & 2560 & 5.44 & 5.74 & 6.36 & 96.3 & 5.44 & 5.76 & 6.34 & 96.2 & 100.0 & 3.0 \\
         & C & 640 & 1.44 & 1.57 & 2.31 & 97.7 & -- & 2.25 & 2.28 & 94.6 & 0.0 & 100.0 \\
         & C & 2560 & 1.44 & 1.43 & 2.41 & 98.8 & -- & 2.00 & 2.41 & 96.7 & 0.0 & 100.0 \\
    \bottomrule
    \end{tabular}}
    \label{tab:j4}
\end{table}

The results match Theorem \ref{thm:asy} and Proposition \ref{prop:test} of the main text, except for the adjusted estimators with $20$ units per cell. In Scenario A with $80$ units per cell, the mean estimated variances of $\tauS$ and $\tauW$ equal four times the arithmetic and harmonic means of the four average cell variances, as predicted by Theorem \ref{thm:asy}(iii). With $80$ units per cell, the Monte Carlo variances agree with their theoretical values within Monte Carlo error. With $20$ units per cell, the Monte Carlo variances of the adjusted estimators are above their theoretical values. In Scenarios A and B, the intervals of $\tauS$ and $\tauW$ maintain their coverage, and in Scenario B the intervals of the fractions over-cover. In Scenario B, the test rejects less often than the nominal rate and is close to its limits, except for the adjusted test with $20$ units per cell, which rejects well above its limit. In Scenario C, the fractions almost never cover $\tau(1)$, and the test almost always rejects. In Scenario A, the test rejects above the nominal rate with $20$ units per cell, more so when adjusted, and the excess shrinks with $80$ units per cell. Thus, with covariates and four cosets, the cells of $\cD$ need more units before the adjusted estimators and the test are close to their limits.

\subsection{Choosing the estimator by the test}\label{supp:sim-pretest}

This study measures the coverage of the pretest estimator, which reports $\tauW$ and its interval when the test of Proposition \ref{prop:test} of the main text does not reject at level $0.05$, and $\tauS$ and its interval when it rejects. The design is the full design $\cD$ of \S\ref{supp:sim-table} with $N=400$, $\eta=0$ and $\kappa$ equal to $0$ or $0.4$. The term $\tau(23)z_2z_3/2$ is added to the potential outcomes, with $\tau(23)$ equal to $0$, $0.1$, $0.2$, $0.3$, $0.4$, $0.5$ or $0.75$, so that Assumption \ref{ass:alias} holds only when $\tau(23)=0$. The potential outcomes are fixed across replications, and each of the $1{,}000$ replications draws a new assignment. Both versions of the estimators are used. Table \ref{tab:pretest} reports the rejection rate of the test and the coverage of the intervals of $\tauS$, of $\tauW$, and of the pretest estimator.

\begin{table}[htbp]
    \caption{Rejection rate (\%) of the test in Proposition \ref{prop:test} of the main text at level $0.05$ and coverage (\%) of the $95\%$ intervals of $\widehat\tau_{\rm S}$, of $\widehat\tau_{\rm W}$ and of the pretest estimator, with $N=400$ over $1{,}000$ replications. The target is $\tau(1)=1$, and $\eta=0$. The pretest estimator reports $\widehat\tau_{\rm W}$ when the test does not reject and $\widehat\tau_{\rm S}$ when it rejects. Assumption \ref{ass:alias} holds only when $\tau(23)=0$. The Monte Carlo standard error of a coverage near $95\%$ is $0.7$ percentage points, and near $88\%$ it is $1.0$ percentage point.}
    \centering
    \begin{tabular}{ccrrrrrrrr}
        \toprule
        & & \multicolumn{4}{c}{Unadjusted} & \multicolumn{4}{c}{Adjusted} \\
        \cmidrule(lr){3-6}\cmidrule(lr){7-10}
        $\kappa$ & $\tau(23)$ & Reject & $\widehat\tau_{\rm S}$ & $\widehat\tau_{\rm W}$ & Pretest & Reject & $\widehat\tau_{\rm S}$ & $\widehat\tau_{\rm W}$ & Pretest \\
        \midrule
        0.0 & 0.00 & 5.4 & 95.9 & 95.9 & 95.8 & 4.6 & 98.8 & 98.7 & 98.7 \\
         & 0.10 & 15.3 & 97.0 & 96.9 & 97.1 & 28.2 & 98.5 & 98.0 & 98.3 \\
         & 0.20 & 47.3 & 97.4 & 97.3 & 97.4 & 73.7 & 98.0 & 97.3 & 98.0 \\
         & 0.30 & 76.8 & 96.6 & 95.8 & 96.4 & 97.3 & 98.1 & 97.7 & 98.0 \\
         & 0.40 & 94.3 & 96.8 & 96.2 & 96.8 & 99.9 & 99.1 & 97.7 & 99.1 \\
         & 0.50 & 99.8 & 97.2 & 95.5 & 97.2 & 100.0 & 98.5 & 96.5 & 98.5 \\
         & 0.75 & 100.0 & 98.1 & 95.8 & 98.1 & 100.0 & 98.2 & 94.0 & 98.2 \\
        0.4 & 0.00 & 4.9 & 97.4 & 97.0 & 96.5 & 5.5 & 99.0 & 99.3 & 98.9 \\
         & 0.10 & 13.5 & 95.2 & 93.8 & 92.1 & 23.7 & 98.6 & 97.1 & 97.1 \\
         & 0.20 & 39.7 & 96.7 & 80.5 & 88.0 & 69.0 & 98.0 & 84.8 & 93.5 \\
         & 0.30 & 69.8 & 97.7 & 58.2 & 87.6 & 95.3 & 97.3 & 61.1 & 95.3 \\
         & 0.40 & 91.6 & 96.8 & 35.7 & 93.0 & 99.8 & 98.7 & 31.8 & 98.5 \\
         & 0.50 & 99.0 & 96.5 & 15.0 & 95.8 & 100.0 & 98.7 & 11.8 & 98.7 \\
         & 0.75 & 100.0 & 96.6 & 1.0 & 96.6 & 100.0 & 98.9 & 0.3 & 98.9 \\
    \bottomrule
    \end{tabular}
    \label{tab:pretest}
\end{table}

The results show that the pretest estimator can under-cover when the cosets have unequal average cell variances, while the interval of $\tauS$ maintains its coverage, as Theorem \ref{thm:asy}(i) of the main text predicts. With $\kappa=0.4$, the limiting weights of $\tauW$ are unequal, so a nonzero $\tau(23)$ makes $\tauW$ biased. For moderate $\tau(23)$, this bias is comparable to the standard deviation of $\tauW$, while the test often fails to reject, and the interval of the pretest estimator under-covers. As $\tau(23)$ grows, the test rejects more often, and the coverage of the pretest estimator returns to that of $\tauS$. The adjusted test has more power, and thus the adjusted pretest estimator loses less coverage. With $\kappa=0$, the limiting weights equal $1/2$, so $\tauW$ is nearly unbiased for every $\tau(23)$, and the pretest estimator maintains its coverage.

\subsection{A sharp bound for the fractions}\label{supp:sim-sharp}

This study checks whether a sharp variance bound could recover the gain of the fractions in Scenario 2 of \S\ref{supp:sim-table}. By Theorem \ref{thm:exact}(i) of the main text, $N\mathrm{var}_j(\widehat\tau_j)=4a_j-S^2\{\ttil_j(k)\}$. The average cell variance $a_j$ is identified because the observed outcomes in each cell are a simple random sample of the potential outcomes of that cell. The heterogeneity $S^2\{\ttil_j(k)\}$, however, is not identified. It depends on how the potential outcomes of the different cells are paired within units, and no unit is observed in more than one cell. The Neyman bound $4a_j$ sets this term to zero. A sharp bound instead subtracts the smallest value of $S^2\{\ttil_j(k)\}$ that is compatible with the distributions of the four cells of the coset. This is the smallest value over all couplings of the four cell distributions, that is, over all ways of reassigning the values of each cell to the $N$ units. For two arms, \citet*{aronow2014sharp} give this bound, and the minimizing coupling sorts the two arms in opposite orders. With four cells, the minimum has no closed form.

To see whether any sharp bound can help, we compute it from the true potential outcomes instead of estimating it from data. The result is the best that a sharp bound could achieve. Write $c_z=g_k(z)/2^{K-p-1}$, so that $\ttil_{j,i}(k)=\sum_{z\in\cC_j}c_zY_i(z)$. We minimize the variance of this sum over the couplings by the rearrangement algorithm, which cycles through the four cells. For each cell, it reorders the signed values $c_zY_i(z)$ of that cell in the order opposite to the sum of the signed values of the other three cells. For the other three cells held fixed, this order gives the smallest variance of the sum. The cycle is repeated until no cell changes, from $20$ random starting couplings, and the smallest variance found is kept. This value is an upper bound on the minimum, so $4a_j$ minus this value is at most the sharp bound. Table \ref{tab:sharp} reports the true heterogeneity, the smallest heterogeneity found, the true variance, the Neyman bound $4a_j$, and the sharp bound computed with the smallest heterogeneity found.

\begin{table}[htbp]
    \caption{Heterogeneity of the coset contrast and $N$ times the variance of the fraction estimator (unadjusted version) in Scenario 2 of \S\ref{supp:sim-table}. The smallest heterogeneity found is over couplings of the cell distributions of the coset and is an upper bound on the minimum. The sharp bound column is $4a_j$ minus this upper bound on the minimum, and is at most the sharp bound.}
    \centering
    \begin{tabular}{crrrrr}
        \toprule
        Coset & $S^2\{\widetilde\tau_j(k)\}$ & Smallest found & $N\mathrm{var}_j(\widehat\tau_j)$ & Neyman bound & Sharp bound \\
        \midrule
        1 & 4.95 & $4.9\times10^{-6}$ & 4.01 & 8.96 & 8.96 \\
        2 & 5.05 & $5.5\times10^{-6}$ & 4.01 & 9.06 & 9.06 \\
    \bottomrule
    \end{tabular}
    \label{tab:sharp}
\end{table}

The results show that the gain of the fractions in Scenario 2 is not recoverable from the cell distributions. For both cosets, a coupling exists under which the coset contrast is nearly constant across units. Thus, the sharp bound essentially equals the Neyman bound, while the true variance is much smaller.

\end{document}